\documentclass[a4paper,fleqn,usenatbib]{mnras}

\usepackage{newtxtext,newtxmath,mathtools,comment}
\usepackage{tablefootnote,nicefrac}
\usepackage[T1]{fontenc}
\usepackage{ae,aecompl}

\def\nodata{--}
\def\sigmaint{$\sigma_{\text{int},i}^2$}
\def\cstat{$C$--stat}
\def\cmin{$C_{\text{min}}$}
\def\chimin{$\chi^2_{\text{min}}$}
\def\Var{\text{Var}}
\def\E{\text{E}}
\def\es{1ES~1553+113}
\def\xmm{\it XMM-Newton\rm}
\def\ovii{O~VII}
\def\spex{\texttt{SPEX}}

\usepackage{graphicx}	
\usepackage{amsmath}	

\def\xmm{\emph{XMM--Newton}}

\title[Systematics with the Cash statistic]{Systematic errors in the maximum--likelihood regression of Poisson count data:
introducing the overdispersed $\chi^2$ distribution}

\author[M. Bonamente]{
  Massimiliano Bonamente$^{1}$\thanks{E-mail: bonamem@uah.edu (MB)}
\\
$^{1}$ Department of Physics and Astronomy, University of Alabama in Huntsville,
Huntsville, AL 35899
}

\date{Accepted XXX. Received YYY; in original form ZZZ}

\pubyear{2022}

\begin{document}
\label{firstpage}
\pagerange{\pageref{firstpage}--\pageref{lastpage}}
\maketitle

\begin{abstract}
This paper presents a new method to estimate systematic errors 
in the maximum--likelihood regression of count data. The method
 is applicable in particular to X--ray
  spectra in situations where the Poisson log--likelihood, or the \emph{Cash} goodness--of--fit statistic, indicate 
  a poor fit
  that is attributable to overispersion of the data. 
  Overdispersion in Poisson data is 
  treated as an intrinsic 
  model variance that can be estimated from
  the best--fit model, using the maximum--likelihood \cmin\ statistic.
  The paper also studies the effects of such systematic errors on the $\Delta C$ likelihood--ratio
  statistic, which can be  used to test for the presence of a nested model component in the regression of
  Poisson count data.
  The paper introduces an \emph{overdispersed $\chi^2$ distribution} that results from the 
  convolution of a $\chi^2$ distribution that models the usual $\Delta C$ statistic, and a zero--mean Gaussian  that
  models the overdispersion in the data. This is proposed as the distribution of choice for the $\Delta C$
  statistic in the presence of systematic errors.
  The methods presented in this paper are applied to 
 \xmm\ data of the quasar \es\ that were used to detect absorption lines from an intervening warm--hot intergalactic medium (WHIM). 
 This case study illustrates how systematic errors
  can be estimated from the data, and their effect on the detection of a nested component, such as an absorption line,
  with the $\Delta C$ statistic.
\end{abstract}




\section{Introduction}

The launch of the first satellites dedicated to
the detection of X--rays beyond the Solar system, 
namely \emph{Uhuru} \citep{giacconi1971}
and the \emph{HEAO} missions \citep{rothschild1979}, marked
the beginning of the field of X--ray astronomy in earnest.
It soon became apparent that the new data provided by these
instruments, typically in the form of the number of counts or
photons as a function of time and energy, required new 
statistical tools for a proper analysis and interpretation.
These new data led to several advances in the study of statistics for count data.

Among the tools needed to interpret the early X--ray data
is the application of the maximum--likelihood method,
devised decades earlier by R.A.~Fisher \citep[e.g.][]{fisher1922,fisher1934},
to these inherently Poisson--distributed integer--count data.
For normally distributed data, the maximum--likelihood method
leads to the $\chi^2$ statistic that features relatively simple
mathematical properties that have long been used by statisticians \citep[e.g.][]{greenwood1996}. 
For Poisson data, X--ray astronomer 
W.~Cash was the first
to show that it is possible to use the maximum--likelihood method
to derive another statistic that is asymptotically 
distributed like a $\chi^2$ distribution. This statistic
is now usually referred as the \emph{Cash} statistic
or \cstat\ \citep{cash1976,cash1979}, and it was further developed
by \cite{cousins1984} and others. In other fields of statistics,
an equivalent Poisson--based log--likelihood is referred to as the \emph{deviance} of the Poisson 
log--likelihood
\citep[e.g.][]{mccullagh1989,cameron2013}, or the \emph{G--squared} statistic \citep{bishop1975}.

The consistent 
use of the $\chi^2$ statistic for integer--count data, in X--ray astronomy
and in related field, is based on the fact that a Poisson distribution
is known to be well approximated by a Gaussian in the limit of a large number of
counts.~\footnote{For example, this approximation is illustrated in Ch.~3 of
\cite{bonamente2022book}.} However, it has now become clear that,
even in the large--count limit, minimization of the $\chi^2$ statistic 
for parameter estimation in models with free parameters will lead to
biased results, when applied to Poisson data 
\cite[e.g.][]{humphrey2009,bonamente2020}.
Such bias is due to the fact that the $\chi^2$ statistic requires
an estimate for the data variance \citep[i.e., according to the \emph{modified $\chi^2$ minimum method}
described by ][pp. 424-434]{cramer1946}, 
and using the number of counts as
its estimate leads to a heavier weight being given to low--count datapoints
in the process of the regression.

On the other hand, the \cstat\ does not have this bias, and it should
be regarded as the statistic of choice for the majority of count data
that are commonplace in X--ray astronomy and related fields. In fact,
the major X--ray fitting packages (\texttt{XSPEC} and \texttt{SPEX})
now give the option
to use either statistic \citep{arnaud1996,kaastra1996}, and \texttt{SPEX} also
provides an estimate of the expected \cstat, under the assumption that the data
follow the best--fit model, based on the approximations of \cite{kaastra2017}.
Nonetheless, the use of the \cstat\ for spectral fitting and other 
like applications
remains hampered by its somewhat more complex mathematics. 
For example, even in the case of a simple linear regression,
a fit to Poisson data does not have the type of simple analytical solution
as in the case of the $\chi^2$ statistic, although it 
was recently shown to have 
a semi--analytical solution that can be easily implemented 
numerically \citep{bonamente2022}. 

In this paper we address an outstanding issue in the fit of
integer--count data with the \cstat, namely how to measure
possible systematic errors that go beyond the
usual Poisson uncertainties. Unlike the case of Gaussian data
fit with a $\chi^2$ statistic, where systematics
can be immediately handled with traditional quadratue addition of 
errors, the Poisson--based \cstat\ offers no such direct modification.
Instead, the problem can be approached with the use of an intrinsic model
variance that reflect the presence of uncertainties in the best--fit model,
instead of additional errors in the data.

This paper is structured
as follows: Sect.~\ref{sec:data} presents the statistical
properties of the data model under consideration, and
Sect.~\ref{sec:intrinsicVariance} presents the new method to address 
systematic errors with the \emph{Cash} statistic, which includes the introduction
of the \emph{overdispersed $\chi^2$ distribution}. Sect.~\ref{sec:es} illustrates the method with
a case study with the \xmm\ data for the source
\es\ recently studied by \cite{nicastro2018} and \cite{spence2023}, and Sect.~\ref{sec:conclusions} presents our conclusions.




\section{Data models for count data}
\label{sec:data}
The data model considered in this paper is $N$ independent measurements
of the type
\begin{equation}
  (x_i, y_i)\; \text{ for } i=1,\dots,N,
  \label{eq:data}
\end{equation}
where $x_i$ is an independent variable assumed to be known exactly (e.g., the wavelength
or the energy of the photons) and $y_i$ is an integer number of counts (e.g., the number of
counts or photons in a given bin). This is the standard model for astronomical X--ray spectra,
but also applies to a variety of data from other fields, and it is often referred to as
\emph{cross--sectional} data. These data are fit to a parametric model $y=f(x)$, where the
function $f(x)$ has $m$ free parameters to be determined according to the maximum--likelihood method.

\subsection{The standard Poisson model}
The simplest assumption
used to estimate
model parameters is that the dependent variable $y_i$ is distributed like a Poisson variable,
\begin{equation}
  y_i \sim \text{Poiss}(\mu_i),
\label{eq:dataPoisson}
  \end{equation}
with an
  unknown parent mean $\mu_i=\mu_i(x_i)$ that is a function of a number of adjustable
  parameter $\theta_j$, $j=1,\dots,m$, and of the independent variable. Common situations 
  in X--ray astronomy are power--law models, 
  or thermal models that are a function of a handful of parameters
  such as temperature and chemical abundances, 
  and more complex models that include absorption lines etc. etc.
  In all cases, the key assumption of this data model 
  is the Poisson distribution of the number of counts that
  are detected in a given \emph{bin}, i.e., a range in the $x$ variable that is
  represented by the characteristic value $x_i$.

  Parameter estimation for these data are obtained from the usual maximum--likelihood method
  pioneered by R.A.~Fisher \citep[e.g., see][]{fisher1934}, in this case
  making use of the Poisson distribution of the data and of the independence among the measurements.
  Under these assumptions, the relevant statistic to be minimized to obtain the best--fit parameters an
  their covariance matrix is the \emph{Cash} statistics or \cstat, defined by
  \begin{equation}
    C = 2 \sum_{i=1}^N \left( \mu_i -y_i +y_i \ln \left( \dfrac{y_i}{\mu_i}\right)  \right).
    \label{eq:cstat}
    \end{equation}
This statistic was developed by \cite{cash1976,cash1979} and \cite{cousins1984}, and it has the convenient 
property that it is asymptotically distributed as a $\chi^2$ distribution for a large number of counts per bin,
\begin{equation}
  C \overset{a}\sim \chi^2(N)
\end{equation}
when the model is fully specified \citep[e.g.,][]{kaastra2017,bonamente2020,bonamente2022book}. 
For a model with $m$ adjustable parameters, the maximum--likelihood regression with the Poisson
statistic leads to the \cmin\ goodness--of--fit statistic,
\begin{equation}
C_{\text{min}}=2 \sum_{i=1}^N \left( \hat{\mu_i} -y_i +y_i \ln \left( \dfrac{y_i}{\hat{\mu_i}}\right)  \right),
    \label{eq:cmin}
    \end{equation}
where $\hat{\mu_i}$ is evaluated for the best--fit parameters $\hat{\theta}_j$. Its
asymptotic distribution for a large number of counts per bin was shown by \cite{mccullagh1986} to be 
approximately
\begin{equation}
  C_{\text{min}} \overset{a}\sim \chi^2(N-m).
  \label{eq:cminAsympt}
\end{equation}
This asymptotic
property is convenient for parameter estimation and for hypothesis/model testing (see, e.g., \citealt{kaastra2017}),
and it
is akin to a similar result that applies to the \chimin\ statistic \citep{cramer1946}.

\subsection{The overdispersed Poisson model}
\label{sec:quasiML}
The greatest restriction imposed by the Poisson distribution is that the variance is equal to the mean,
\begin{equation}
\Var(y_i) = \mu_i,
  \label{eq:VarPoisson}
\end{equation}
where $\mu_i$ is the parent mean of the data--generating process for the $i$--th bin.
A common occurrence in datasets across the sciences, including astronomy, 
is that the measured data are \emph{overdispersed} relative to the
simple assumption on \eqref{eq:VarPoisson}, e.g., see discussion in Chapter~3 of \cite{cameron2013}.
Reasons for this larger--than--${\mu_i}$ variance in the data include the inadequacy of the model --- i.e., more
explanatory variables may be required for multi--variable regression, or better parameterization of the model ---
or an intrinsic model variability that makes each independent measurement $y_i$ have a larger--than--Poisson variance.

Regardless of origin, overdispersion can be modelled via a function 
\begin{equation}
\Var(y_i)=\omega_i=\omega(\mu_i,\alpha)
\end{equation}
that introduces an additional degree of freedom in the form of the parameter $\alpha$. While in principle the function can have any form,
two convenient parameterizations proposed by \cite{cameron2013} are
\begin{equation}
  \begin{cases}
    \begin{aligned}
      \omega_i&=(1+\alpha)\mu_i\;& \text{ (NB1 model)} \\
      \omega_i&=\mu_i+\alpha \mu_i^2\;& \text{ (NB2 model)}
    \end{aligned}
  \end{cases}
  \label{eq:VarNB}
\end{equation}
where the model names derive from their use in the negative binomial regression by \cite{cameron1986}. This
paper does not make use of the negative binomial distribution, which is featured  in an alternative method of regression
for count data \citep[e.g., see][]{hilbe2011,hilbe2014}, yet this parameterization 
remains applicable. When $\alpha=0$, the model returns the usual
Poisson regression. When $\alpha >0$, on the other hand, the data no longer follows the Poisson distribution. In principle, 
it is also possible to have $\alpha < 0$, which corresponds to \emph{underdispersed} data. This case, however, occurs less
often in practice, and \eqref{eq:VarNB} are normally used for $\alpha>0$.  An equivalent parameterization for the NB1 model
that is common in the generalized linear model (GLM) literature \citep[e.g.][]{mccullagh1989} is to set
\begin{equation}
  \phi=(1+\alpha)
  \label{eq:phi}
\end{equation}
with $\phi>1$ indicating overdispersed data.

The choice to allow overdispersion in the data means that minimization of the statistic described by 
\eqref{eq:cstat} no longer represents a maximum--likelihood condition, since the data
are no longer Poisson--distributed. Fortunately, 
there are situations when the usual Poisson regression, with best--fit parameters $\hat{\theta}_j$ 
obtained from the minimization of \eqref{eq:cstat}, continue to apply to the overdispersed Poisson model.
In particular, \cite{gourieroux1984} have shown that, for a family of exponential distributions that includes 
the Poisson, the best--fit parameters estimated via the usual likelihood remain \emph{consistent} even when
the distribution is misspecified, provided that the mean is correctly specified. In this case,
the method of estimation is referred to as a \emph{quasi maximum likelihood}.~\footnote{ This method
is also referred to as \emph{pseudo maximum likelihood} in the statistical literature.}

In practice, these results let us continue with the usual Poisson likelihood under the assumption that
the Poisson mean is  correct. The problem then turns to the estimate of the overdispersion parameter $\phi$
defined in \eqref{eq:phi}. The best--fit parameters from the Poisson quasi--ML
method are asymptotycally normally distributed, with a variance that clearly depends on $\phi$ \citep[see, e.g.,][]{cameron2013}. 
%
For the data analyst, what is of most interest is an estimate of the degree of overdispersion in the count data.
The standard estimators for $\phi$ and $\alpha$ in the two models for the data variance in \eqref{eq:VarNB} are
\begin{equation}
  \begin{cases}
  \begin{aligned}
    \hat{\phi}_{\text{NB1}}&=\dfrac{1}{N-m} \sum_{i=1}^N \dfrac{(y_i-\hat{\mu_i})^2}{\hat{\mu_i}} \\
    \hat{\alpha}_{\text{NB2}}&=\dfrac{1}{N-m} \sum_{i=1}^N \dfrac{(y_i-\hat{\mu_i})^2 -\hat{\mu_i}}{\hat{\mu_i}^2}.
  \end{aligned}
  \end{cases}
  \label{eq:phiHat}
\end{equation}
Justification for these estimators is provided in \cite{cameron2013,gourieroux1984,gourieroux1984b}, although 
other estimators are also available \citep[see, e.g.,][]{cameron1986, dean1989,dean1992}.
The two equations in \eqref{eq:phiHat} provide  point estimates for the overdispersion in count data, according to the
alternative parameterizations of \eqref{eq:VarNB}. Uncertainties in these estimates can be provided by standard
error--propagation methods (also known as the \emph{delta} method). A value $\hat{\phi}_{\text{NB1}}>1$ 
or $\hat{\alpha}_{\text{NB2}}>0$ 
indicates overdispersion in the data.

\subsection{The Gaussian model}
For X--ray spectral analysis, the use of the $\chi^2$ statistic remains pervasive, even for Poisson--distributed data. 
The $\chi^2$ goodness--of--fit statistic results from the use of
an alternative data model, whereby the same data as \eqref{eq:data} presume a parent distribution
\begin{equation}
  y_i \sim \text{Gauss}(\mu_i,\sigma^2_i)
\label{eq:dataGauss}
\end{equation}
where a \emph{parent} variance $\sigma^2_i$ is also required, 
in addition to the usual parent mean $\mu_i$. 
The widespread use of the $\chi^2$ fit statistic for integer--valued count data
is primarily due to its ease of use and interpretation. This includes a reduction--of--degrees--of--freedom theorem
established by H.~Cramer \citep{cramer1946}, which 
establishes the distribution of the $\chi^2_{\text{min}}$ statistic as
\begin{equation}
  \chi^2_{\text{min}} \sim \chi^2(N-m),
  \label{eq:cramer}
\end{equation}
where $m$ is the number of free parameters in the model and $N$ the usual number of independent
Gaussian--distributed data points. This result holds under rather general conditions that are usually satisfied by X--ray
astronomical data.~\footnote{This theorem is 
also discussed in Sect.~12 of \cite{bonamente2022book}.}

The parent 
variance in \eqref{eq:dataGauss} is usually approximated via the \emph{data} variance, 
typically the number of counts
in the bin,  $\hat{\sigma}_i^2= {n_i}$ \citep[see, e.g.][]{bevington2003,bonamente2022book}. This 
approximation follows the \cite{cramer1946} modified minimum $\chi^2$ criterion,
whereby the asymptotic distribution \eqref{eq:cramer} holds under the assumption of a \emph{fixed} data variance.
It has been documented in the astronomical statistical literature that,
even in the large--count limit where a Poisson distribution is in fact well approximated by the normal distribution,
the use of the $\chi^2$ distribution leads to biased results, 
compared to the use of the \cstat\ \citep[e.g.][]{humphrey2009,bonamente2020}. This is primarily due to the 
approximation of the variance in the data model \eqref{eq:dataGauss} with a number that
is smaller for bins with lower count--rates, which unduly carries a larger weight in the regression.

\subsection{Regression with other distributions}
Another possibility
is to assume that the data follow an alternative distribution, 
such as  the \emph{Conway--Maxwell--Poisson} distribution
\citep[e.g.][]{conway1962,shmueli2005,sellers2010}, the generalized Poisson distribution
\citep[e.g.][]{consul1973,famoye1993},  or the negative binomial distribution \citep[e.g.][]{hilbe2011,hilbe2014}.
These distributions introduce additional parameters that can conveniently model the data variance and thus
allow for over-- or under--dispersed data.
Such modifications, however, come at a cost in terms of ease of use and interpretation,  
including the identification of an alternative goodness--of--fit statistic in place of the Poisson--based deviance or \cstat.

It is often
preferrable to retain the Poisson distribution for the analysis of astronomical data, given its 
relative ease of use and the fact that 
most datasets are believed to be derived from a Poisson process with a fixed rate $\mu_i$. This paper
therefore continuess with the assumption that the data are Poisson--distributed, and develop a
new method that accounts for overdispersion.

\section{Systematic errors in the maximum--likelihood regression of Poisson--distributed 
count data}
\label{sec:intrinsicVariance}
It is traditional in X--ray astronomy and related fields to consider two
separate types of uncertainties in the measurement of the $y_i$ data. The first is 
a so--called \emph{statistical error}, typically used to denote the uncertainty associated
with the photon--counting experiment (e.g., the square root of the number of counts).
All other sources of uncertainty 
are often referred to as \emph{systematic errors}, although they may 
 well be errors that are random in nature in the same way as the photon--counting process itself.
Examples of the latter type of errors are uncertainties associated with other aspects of the photon--collection
process, such as uncertainties associated with the calibration of the instrument, or with other aspects of the
analysis. What generally distinguishes the first from the second type of error is that the first are inherent in the
collection process (i.e., in the Poisson distribution that underlies the $y_i$ variable), while the second can often
be reduced by a careful reduction and analysis of the data. Although this distinction is somewhat arbitrary,
it will be used in the remainder of this paper.

The usual method to address the presence of additional sources of systematic errors
with the $\chi^2$ statistic for Gaussian data is to modify the variance
 in \eqref{eq:dataGauss} until
the fit statistic $\chi^2$ becomes `\emph{acceptable}'. This method leverages the fact that
the variances $\sigma^2_i$ are specifically part of the data model, per \eqref{eq:data} and \eqref{eq:dataGauss}.
A review of these methods is provided, for example,
in Ch.~17 of \cite{bonamente2022book}, or in the textbook by \cite{bevington2003}.
Prior to continuing the discussion of systematic errors, it is necessary to emphasize the meaning of the
word \emph{acceptable} when used in conjunction with hypothesis testing. Hypothesis testing is a process
whose outcome is that of either discarding a null hypothesis at a given level of significance, or 
merely failing to discard it (see, e.g., the recent  statemente by the
American Statistical Association on $p$ values and hypothesis testing, \citealt{asa2016}).
It is nonetheless reasonable to say that a hypothesis is \emph{acceptable} when one may not reasonably discard it (say, at
the 3 or 5 $\sigma$ level of confidence), provided
it is understood that the hypothesis may never be conclusively \emph{accepted} as the only possible model for the data.

For the type of Poisson data that are of common occurrence in X--ray astronomy, i.e., data following the
assumption \eqref{eq:dataPoisson} and the resulting \cstat\ fit statistic \eqref{eq:cstat}, this direct avenue is not
possible. The reason is that the variance of a Poisson data point is equal to its mean, and one may not 
independently specify both, as one can for the Gaussian distribution and the associated $\chi^2$ statistic
according to \eqref{eq:dataGauss}. A convenient workaround that is well established in the
statistical literature is the overdispersed Poisson regression that was discussed in Sect.~\ref{sec:quasiML}.
This quasi--ML method provides a convenient means to retain the Poisson--based maximum--likelihood
best--fit statistic (i.e., the deviance or \cstat),
while providing an estimate for the degree of overdispersion.

Within this statistical framework, this paper presents a new method to address the presence of overdisperion in count data
and its effects on the \emph{Cash} statistic. The method is based on the
interpretation of the overdispersion as an \emph{intrinsic
model variance} that is modelled by an appropriate normal distribution of zero mean. Following
this assumption, this additional
source of uncertainty in the model is folded with the distribution for the usual Poisson--based 
\cstat, resulting in a
modified distribution that can be used for the hypothesis testing of regression with overdispersed 
Poisson data. This method is described in the remainder of this section.

\subsection{Measurement of the intrinsic model variance with the \emph{Cash} statistic}
\label{sec:Cmin}
The means to estimate systematic errors in Poisson--distributed data that 
are fit to a parametric model, as is
customary in X--ray astronomy, is provided 
by a change in perspective with regards to the roles played in the regression by the data and  the model.
Specifically, instead of requring that the data points 
$y_i$ have a larger variance than what is estimated by their Poisson distribution,
it is reasonable to assume that the model has an \emph{intrinsic variance}. 
In practice, this means assuming that the parent model
$\mu_i$ of a given datum $y_i$ is not a fixed (and unknown) number, as implied by its Greek--letter notation,
but it follows a distribution with a given variance, or
\begin{equation}
  \mu_i \sim \text{Gauss}(\mu_{i0},\sigma_{\text{int},i}^2),
  \label{eq:modelVariance}
\end{equation}
where $\mu_{i0}$ is a parent value,
\sigmaint\ is the intrinsic variance, and $i$ denotes the bin.
This assumption is akin to what is done in the case of systematic errors for normal data -- i.e., when the variance of the data
are increased -- except that the additional source of variance is now associated with the model, and not the data. 
Of course the  intrinsic variance is not known a priori, but must be estimated from the data. This is similar
to the case of the overdispersed Poisson model (see Sect.~\ref{sec:quasiML}) which requires
an estimate for the overdispersion parameters $\phi$ or $\alpha$.

To estimate the intrinsic variance in the model, it must be assumed that the best--fit model is an acceptable
description of the data. In statistical terms, this \emph{null hypothesis} can be rephrased as the requirement that 
the data are compatible with being drawn from the parent model, as measured from the fit statistic of choice,
and at a given level of confidence.
 Given that the fit statistic of choice for Poisson data is \cmin\ according the \eqref{eq:cmin},
the model acceptability can be stated in terms of the requirement that the measured \cmin\ 
is statistically consistent with its
parent distribution, under the null hypothesis. In the asymptotic limit of a
large number of counts per bin, and regardless of the number of degrees of freedom,
\cite{mccullagh1986} showed that 
the statistic converges towards a $\chi^2$ distribution, namely 
\begin{equation}
  C_{\text{min}} \overset{a}{\sim} \chi^2(\nu)
\label{eq:cminNorm}
\end{equation}
where $\nu=N-m$ is the number of degrees of freedom. The proof provided by \cite{mccullagh1986} assumes that the
model is linear (or, more precisely, log--linear), and it is based on the calculation
of conditional cumulants \citep{mccullagh1984}, and therefore the extrapolation to non--linear models
such as the one used in \cite{spence2023} would require further theoretical justification.
To the best of the author's knowlegde, to date a general proof of \eqref{eq:cminNorm} 
for a general non--linear model has not been obtained, but it will be hereafter assumed as a first--order approximation
for a general non--linear model.

Another important caveat is that, in the low--count regime, \eqref{eq:cminNorm}
will in general not apply, and additional considerations need to be used. A discussion of the distribution
of \cmin\ in the low--count and few--bin regimes is also provided in \cite{mccullagh1986}, and
also in \cite{bonamente2020} and \cite{kaastra2017}.
Qualitatively speaking, \eqref{eq:cminNorm} is understood with the convergence of the Poisson distribution to
a normal distribution in the large--count limit. For clarity, the domains of applicability of the 
results provided in this paper will be  summarized in Sect.~\ref{sec:domains}.

\subsubsection{The asymptotic distribution of \cmin\ with systematic errors}
\label{sec:cminSys}
The problem now turns to the determination of the effect of the intrinsic model variance \eqref{eq:modelVariance}
on the asymptotic distribution of \cmin, obtained under the assumption of Poisson measurements.
For this purpose, we make the following approximations, which correspond to common
experimental conditions:

(a) For a large value of the number of degrees of freedom, the $\chi^2(\nu)$ distribution is
equivalent to a $N(\nu, 2\nu)$ normal distribution, especially for large values of the statistic
that are not affected by the positive--definite nature of the $\chi^2$ distribution. 
This is consistent with \eqref{eq:cminNorm}, for large $\nu$.
The case of a small $\nu$ is addressed in Sect.~\ref{sec:DeltaC}, which describes
the effect of intrinsic variance on the 
$\Delta C$ statistic.

(b) We assume small values for the intrinsic variance, $\sigma_{\text{int},i}/\mu_{i0} \ll 1$, meaning
that there is a small \emph{fractional} systematic error. Accordingly,  
the method uses a simple error--propagation or delta method to estimate the additional variance 
to the \cmin\ statistic introduced by the assumption \eqref{eq:modelVariance}. 

(c) Following the two earlier assumption, 
the systematic error (or, more precisely, the intrinsic variance) according to \eqref{eq:modelVariance} leads to
a $Z=C_{\mathrm{min}}$ statistic that can be thought of as
the sum of two independent variables, $Z=X+Y$, with
\begin{equation}
 \begin{cases}
   X \sim \chi^2(\nu) \overset{a} \sim N(\nu, 2\nu)\\
   Y \sim N(0,\hat{\sigma}^2_C),
 \end{cases}
  \label{eq:distr}
\end{equation}
with $\nu=N-m$. The null mean for $Y$ means that
the systematic error does not provide a net mean value or bias to the statistic, and $\hat{\sigma}^2_C$
represents an additional variance term of \cmin\ associated with the systematic errors, which will
be estimated from the data.
This ensures that the modified distribution for the \cmin\ statistic remains approximately normal,
with increased variance relative to the standard Poisson regression.

\subsubsection{The estimation of the intrinsic variance}
\label{sec:intrinsicVariance2}
The model described in the previous section has introduced two quantities, 
the intrinsic model variance $\sigma_{\text{int},i}^2$
and a `design' variance of the \cmin\ statistic $\hat{\sigma}^2_C$ that is required for acceptability
of the null hypothesis. This section addresses their estimates.

For a given application, the first step is the estimate of the required intrinsic variance
for the fit statistic, which is denoted by
\begin{equation}
  \Var_i(C_{\text{min}}) = \hat{\sigma}^2_C.
  \label{eq:VarCi}
\end{equation}
The meaning of \eqref{eq:VarCi} is that the \cmin\ statistic requires an additional 
variance, in a given amount of $\hat{\sigma}^2_C$,
in order for the measured statistic to be consistent with its expectation
under the null hypothsis.
According to \eqref{eq:distr}, the total variance of the \cmin\ statistic is then
\begin{equation}
\Var(C_{\text{min}}) = 2 \nu + \hat{\sigma}^2_C,
  \label{eq:VarTotal}
\end{equation}
with the assumption that the \cmin\ statistic retains the normal distribution, in the 
asymptotic limits of a large number of counts per bin, and a large number of bins.

The problem now turns to the evaluation of the left--hand term of \eqref{eq:VarCi},
and its relationship to the sought--after intrinsic model variance \sigmaint\ in \eqref{eq:modelVariance}.
The intrinsic model variance can be evaluated following its definition of
\begin{equation} \begin{aligned}
  \Var_i(C_{\text{min}}) & =   \E\left[\left(C_{\text{min}}(\mu_i)- \E\left[C_{\text{min}}(\mu_i)\right]\right)^2\right] \\
 & \simeq \E\left[\left(C_{\text{min}}(\mu_i)-C_{\text{min}}(\hat{\mu_i})\right)^2\right],
\end{aligned}
\label{eq:VariCmin}
\end{equation}
with the approximation of $\E\left[C_{\text{min}}(\mu_i)\right] \simeq C_{\text{min}}(\hat{\mu_i})$.
The meaning of \eqref{eq:VariCmin} is  that $\mu_i$ is a considered a random variable, distributed according to
\eqref{eq:modelVariance}, while $\hat{\mu_i}$ is the usual maximum--likelihood estimate. 
In other words, this variance is calculated with respect to the distribution
\eqref{eq:modelVariance}, and \emph{not} with respect to the distribution of the data
according to \eqref{eq:dataPoisson} --- this is the reason for
the $i$ subscript in the notation of the variance of \eqref{eq:VarCi} and \eqref{eq:VariCmin}. 
In fact,  the Poisson distribution for the
data $y_i$ is  responsible for the 
variance of \cmin\ that was already calculated, i.e., using the 
large--count approximation as in \eqref{eq:cminNorm}, 
or according to the methods of \cite{kaastra2017} for the low--count regime.

To evaluate \eqref{eq:VariCmin}, use the \cstat\ defined in \eqref{eq:cstat}, 
\[ 
\begin{aligned}
  C_{\text{min}}(\mu_i)- \E\left[C_{\text{min}}\right] = 
	2 \left(\sum_{i=1}^N (\hat{\mu}_i - \mu_i) + y_i \ln \left(\dfrac{\mu_i}{\hat{\mu}_i}\right)\right) \\
  \simeq 2 \left(\sum_{i=1}^N  (\hat{\mu}_i - \mu_i) + y_i\left(\dfrac{\mu_i-\hat{\mu}_i}{\hat{\mu_i}}\right) \right) 
	= 2 \left(\sum_{i=1}^N  (  \mu_i-\hat{\mu}_i) \left(\dfrac{y_i}{\hat{\mu}_i} -1\right) \right)
\end{aligned}
	\]
where it was assumed a small value for the fractional
intrinsic error, $\sigma_{\text{int},i}/\mu_i$, so that only the first--order term in the Taylor series
expansion of the logarithm was retained.
With this approximation, 
\eqref{eq:VariCmin}  leads to
\begin{equation} \Var_i(C_{\text{min}}) \simeq
\E \left[ \left(2 \sum_{i=1}^N (\mu_i-\hat{\mu_i})
\underbracket{\left( \dfrac{y_i}{\hat{\mu_i}}-1\right)}_{ (a)} \right)^2\right],
  \label{eq:bracket}
\end{equation}

where the term $(a)$ highlighted by the square bracket is independent of $\mu_i$.
Moreover, upon evaluation of the square of the sum, the cross--product terms have null
expectation, under the assumption that the $\mu_i$ terms are uncorrelated. This leads to
the simple result that
\begin{equation} \Var_i(C_{\text{min}}) =  \hat{\sigma}^2_C  \simeq
4 \left( \sum_{i=1}^N  \sigma_{\text{int},i}^2 \left( \dfrac{y_i}{\hat{\mu_i}}-1\right)^2 \right),
  \label{eq:1}
\end{equation}
with the intrinsic variance of the model being 
\[ \sigma_{\text{int},i}^2 = \E [(\mu_i-\hat{\mu_i})^2].
\]
The right--hand term of \eqref{eq:1} can be further approximated by the assumption
that $y_i \sim \text{Poiss}(\hat{\mu_i})$, i.e., 
it is true that, on average 
\[(\hat{\mu_i}-y_i)^2 \simeq \hat{\mu_i} \simeq y_i.\] 
Further assuming that the intrinsic fractional error is the same for
all datapoints, i.e., $\sigma_{\text{int},i}/\hat{\mu_i}$ is a constant, we obtain
the final result
\begin{equation}
  \dfrac{\sigma_{\text{int},i}}{\hat{\mu_i}} \simeq \dfrac{1}{2}
    \dfrac{\hat{\sigma}_C}{\sqrt{\sum_{i=1}^N y_i}}.
    \label{eq:sigmai}
\end{equation}
Equation~\ref{eq:sigmai} is the sought--after result, namely 
the approximation that relates the fractional
intrinsic scatter of the model (${\sigma_{\text{int},i}}/{\hat{\mu_i}}$) to the design value $\hat{\sigma}^2_C$ 
required for consistency between the measured \cmin\ and 
its parent distribution under the null hypothesis.~\footnote{Notice that, 
in this estimate of the variance, the right--hand
term $(y_i/\mu_i-1)$ was considered constant with respect to the intrinsic variance defined by 
\eqref{eq:modelVariance}. 
The variance of \cmin\ with respect to $y_i$ and $\hat{\mu_i}(y_i)$ 
is calculated separately as the 'proper' variance of \cmin\ according to the
Poisson disrtibution of the data, as was also indicated earlier in the paper.}
This approximation provides a simple means to estimate an average fractional or
percent intrinsic variance of the model that is required to assure an acceptable fit.

\subsection{The intrinsic model variance 
for hypothesis testing of a nested component with the $\Delta C$ statistic}
\label{sec:DeltaC}
The intrinsic model variance estimated from the \cmin\ statistic according to \eqref{eq:sigmai}
makes it such the expected statistic, under the null hypothesis, is now consistent with the
measured value. In fact, the method of estimation started with the
requirement \eqref{eq:VarCi}, designed exactly to achieve such agreement. Sect.~\ref{sec:es}
will present a detailed case study of how such estimate is performed in practice.
In other words, the intrinsic model variance \emph{assumes} that the fit is acceptable,
and therefore hypothesis testing with \cmin\ is no longer meaningful.

Another common use of the \cmin\ statistic, also shared by the \chimin\ statistic,
is when the statistic
\begin{equation}
\Delta C = C - C_{\text{min}}
 \label{eq:DeltaCDef}
\end{equation}
is used to test the significance
of a nested model component. This method was originally developed by \cite{lampton1976} 
for the $\Delta \chi^2$ statistic,
and it is based on the theory presented in \cite{cramer1946}. A description of
the $\Delta C$ statistic is presented in Ch.~16 of \cite{bonamente2022book}, with the 
main result of
\begin{equation}
  \Delta C \overset{a}{\sim} \chi^2(l) 
  \label{eq:DeltaC}
\end{equation}
in the asymptotic limit of a large number of measurements, and with $l$ interesting (or free)
parameters for the nested component.

The use of the $\Delta C$ statistic  consists of calculating the \cmin\ fit statistic for the full model that includes
a nested component, and the calculating the $C$ statistic with the nested model component
zeroed--out (referred to as the reduced model), so that clearly $C \geq C_{\text{min}}$ and $\Delta C \geq 0$.
In this case, the best--fit model
$\hat{\mu_i}$ has $l$ additional  parameters associated with its
nested component, with a model value of $\mu_i^{'}$ when the nested component is zeroed out. This is the
case, for example, for  the data of Tables~5 and~6 in \cite{spence2023}, where
the $\Delta C$ values correspond to setting to zero 
  a spectral line component that has $l$=1 or 2 additional free parameters, relative to a model for the continuum
 emission of a bright X--ray quasar. These data will be described and analyzed
 in more detail in the case study of Sect.~\ref{sec:es}. 

As a likelihood--ratio statistic, $\Delta C$ must obey a number
 of conditions in order to feature the asymptotic distribution \eqref{eq:DeltaC}.
 These conditions are conveniently described by \cite{protassov2002}, including the requirement that
 the null value of the nested component must not be at the boundary of the allowed parameter space.

\subsubsection{The intrinsic variance for the $\Delta C$ statistic}
Qualitatively speaking, if the contribution from the nested model component is within
the value of the intrinsic error estimated by \eqref{eq:sigmai}, 
one would expect that it is not possible to determine whether the nested component
is needed or not.  For example, an absorption line--like feature that is within 10\% of the continuum,
is not expected to be detected conclusively 
in the presence of a $\pm 10$\% intrinsic error.

To investigate this problem quantitatively, 
consider the deviation between
the the reduced or baseline model $\mu_i^{'}$ and the best--fit full model,
\begin{equation}
  \delta_i = \mu_i^{'} -\hat{\mu}_i.
    \label{eq:deltai}
\end{equation}
This equation applies only to
a few bins where the nested model component is significant, say $i=1,\dots, n,$ 
typically with $n \ll N$,
with $\delta_i =0$ in all other bins. 
This is the case for a small--scale fluctuation in the form of a nested absorption or 
emission line--type component in a restricted wavelength range,
superimposed on a baseline continuum model, often 
featured in the detection of absorption lines \citep[e.g.][]{nicastro2018,spence2023}.
In \eqref{eq:deltai}, it is assumed that the distribution of the deviation
\[
  \delta_i \sim N(0, \sigma^2_{\text{int},i})
\]
assuming that the baseline model $\mu_i^{'}$
is a random variable with the same intrinsic variance as the full model $\mu_i$, see \eqref{eq:modelVariance}.
A value of zero for its mean also represents
the null hypothesis that the nested component is not present.

The $\Delta C$ statistic can therefore be approximated as
\[ \Delta C \simeq 2  
\sum_{i=1}^n \delta_i \left( 1- \dfrac{y_i}{\hat{\mu_i}}\right),\]
following the same assumption of small intrinsic variance  as in Sect.~\ref{sec:intrinsicVariance2}.
 Also,
the term in parentheses in the right--hand side is 
independent of the intrinsic variance, same as 
for the variance of \cmin\ in \eqref{eq:bracket}. 
It follows that the variance of $\Delta C$ associated with the intrinsic model variance
is approximated as
\begin{equation}
  \Var_i(\Delta C) = \hat{\sigma}^2_{\Delta C}  \simeq 4 \left( \dfrac{\sigma_{\text{int},i}}{\hat{\mu_i}} \right)^2 \sum_{i=1}^n \hat{\mu_i} 
  \simeq 4 \left( \dfrac{\sigma_{\text{int},i}}{\hat{\mu_i}} \right)^2 \sum_{i=1}^n y_i. 
    \label{eq:VarDeltaC}
\end{equation}
This result  is identical to the equation for $\Var_i(C_{\text{min}})$ 
that yielded \eqref{eq:sigmai}, except that the sum extends 
only over the few bins ($n \ll N$) where the nested model has an effect. 
Equation \ref{eq:VarDeltaC} is the sought--after value for the
intrinsic variance of $\Delta C$, using the intrinsic variance in the data according to \eqref{eq:sigmai}.

\subsubsection{The distribution of the $\Delta C$ statistic in the presence of 
intrinsic variance}
\label{sec:DeltaCDistr}
The analysis now turns to the question of the distribution of $\Delta C$ in the small--$\nu$ limit, in the presence
of an intrinsic model variance. In this section, $\nu$ denotes the number of free parameters
in the nested component, which is also the number of 
degrees of freedom
of the associated asymptotic $\chi^2$ distribution, whereas in the previous section it was used to denote the
number of degrees of freedom for the entire regression.

\begin{figure}
  \includegraphics[width=3.5in]{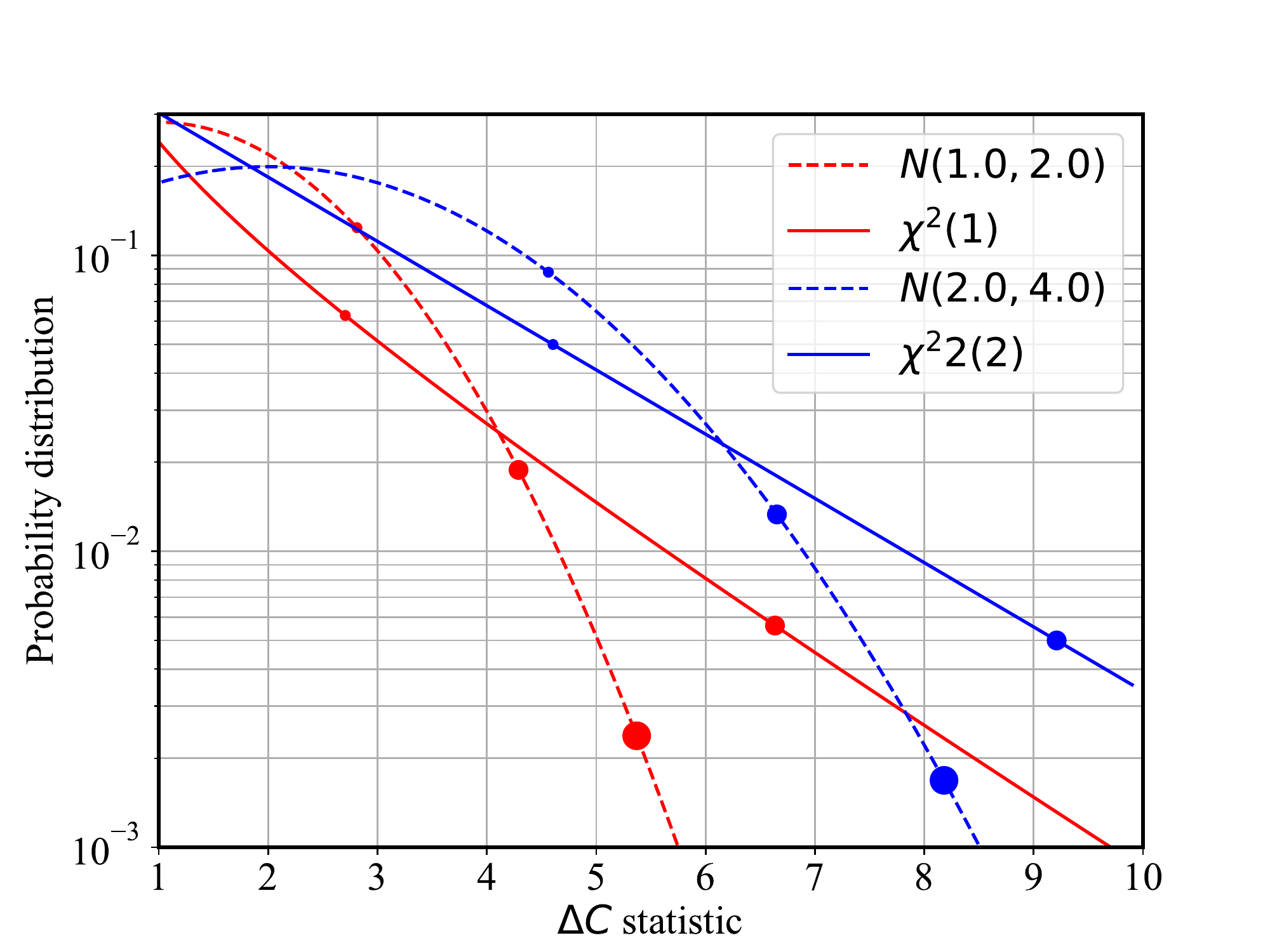}
  \caption{Comparison of $\chi^2$ and Gaussian distributions for selected numbers of degrees
  of freedom. In increasing order of marker size are also reported the $p=0.9, 0.99, 0.999$
  one--sided critical values of the distributions.}
  \label{fig:normChi}
\end{figure}

Under the assumption of a large number of measurements,
the sampling distribution of the $\Delta C$ statistic
is approximately a $\chi^2(\nu)$ distribution,
with a mean of
$\nu$ and variance of $2\nu$. Given the small number of free parameters usually employed in
the $\Delta C$ statistic for a nested model component, say $\nu=l=1$ or 2 (e.g., one or two extra parameters
in the nested component),
the approximation of the $\chi^2(\nu)$ distribution with a normal distribution is poor.
Figure~\ref{fig:normChi} illustrates the heavier right tail of the $\chi^2(\nu)$ distribution for $\nu=1$ and 2,
compared to the corresponding Gaussian $N(\nu,2\nu)$, which leads to asymptotically
larger critical values for the $\chi^2(\nu)$ distribution, as $p$ increases.
For $p=0.9$, or a 90\% one--sided confidence interval, the critical values of the
two distributions are similar, and for larger $p$--values the Gaussian approximation
systematically underestimates them. It is clear that, in general, it is not
advisable to employ the normal approximation for the $\chi^2(\nu)$ distribution, as was done for the
study of the \cmin\ distribution when $\nu$ is a large number.

For the $\Delta C$ statistic, the distributions in \eqref{eq:distr} are modified so that
its parent $B$ distribution is assumed to be the sum $B=X+Y$ of two variables with
\begin{equation}
 \begin{cases}
   X   \sim \chi^2(\nu)\\
   Y \sim N(0,\hat{\sigma}^2_{\Delta C}),
 \end{cases}
  \label{eq:distrDeltaC}
\end{equation}
where $\nu=l$ is the number of free parameters of the additional nested component.
The notation is such that $B$ is an \emph{overdispersed $\chi^2$ distribution} that becomes the usual $\chi^2(\nu)$ 
distribution when the additional
variance $\hat{\sigma}^2_{\Delta C}$, induced by the presence of systematic errors according to \eqref{eq:modelVariance}, is zero.

There are two possible avenues for the estimate of critical values and other parameters of the overdispersed $\Delta C$
statistic according to \eqref{eq:distrDeltaC},
when $\nu$ is a small number of degrees of freedom.
The first one makes a crude approximation of the $\chi^2(\nu)$ distribution with a normal distribution which,
as was already warned, 
should be used with caution, and it is described primarily for the illustrative purposes below in Sect.~\ref{sec:normApprox}. 
The second uses a convolution of the two distributions in  \eqref{eq:distrDeltaC} --- the  $\chi^2$ distribution 
and a normal distribution of zero mean --- and it is  introduced in Sect.~\ref{sec:modChi2} 
as the \emph{overdispersed $\chi^2$ distribution}. This  should be regarded as
the method of choice for virtually all applications.

\subsubsection{Normal approximation to the $\chi^2(\nu)$ distribution}
\label{sec:normApprox}
This method is  completely equivalent to that of Sect.~\ref{sec:Cmin} above, and it makes use of
the normal approximation to the parent $\chi^2(\nu)$ distribution, and uses the linear addition
of variances in \eqref{eq:VarTotal} to determine the critical value of the $\Delta C$ distribution that accounts for
the systematic error. Given that $\nu$ is typically a small number, the results of this method are approximate, and
it is generally preferrable to use the overdispersed $\chi^2$ distribution described below. Nonetheless,
given its simplicity and for the sake of illustration, it is worthwhile to discuss it further.

As shown in Figure~\ref{fig:normChi}, the critical values of the $\chi^2(\nu)$ distribution
are larger than those of the matching $N(\nu,2 \nu)$ distribution, especially for small residual probabilities, given the
heavier tail of the $\chi^2(\nu)$ distribution.
This method is therefore approximate and
conservative, in that
a null hypothesis would never be erroneously deemed \emph{acceptable} (i.e., the additional component
\emph{is} warranted) at the $p$ confidence level, if it truly wasn't according to the asymptotic $\chi^2$ distribution.
On the other hand,
the rejection of the null hypothesis (i.e., the additional component \emph{is not} warranted)
at the $p$ confidence level may be incorrectly determined. 

When using this approximation,
one should state that the null hypothesis was rejected at a confidence level $\leq p$, where $p$
is the chosen confidence level for the test.
The advantage of this approximation is that one can immediately use the addition of variances
in  \eqref{eq:VarTotal} and the Gaussian distribution function to calculate critical values. Further,
Figure~\ref{fig:normChi} indicates that the 90\% critical values of the two distributions ($\chi^2(\nu)$
and $N(\nu,2\nu)$) are nearly identical for $\nu=1, 2$, suggesting that a choice of $p=0.9$ is likely to minimize
the errors associated with this approximation. Given that a more accurate method is described below,
this approximation is not studied further, and it is generally not recommended for most applications.

\subsubsection{The overdispersed $\chi^2$ distribution}
\label{sec:modChi2}
 A more accurate method to study the distribution of the $\Delta C$ statistic according to
 the model of \eqref{eq:distrDeltaC} consists of retaining the $\chi^2(\nu)$ distribution for 
 the Poisson contribution to the $\Delta C$ statistic.
This means that, while
the variances will still add linearly according to \eqref{eq:VarTotal} thanks to the assumption
of independence of the contributing random variables,
 the distribution of $\Delta C$ according to  \eqref{eq:distrDeltaC}
 is \emph{not} Gaussian. In this case, the parent distribution of the $\Delta C$ statistic becomes the \emph{convolution}
 of the two distributions,
 \begin{equation}
   f_B(z; \nu,\sigma^2)=\int_{-\infty}^{+\infty} f_{\chi^2}(z-y; \nu) f_{N}(y; 0,\sigma^2)\,dy
   \label{eq:conv}
 \end{equation}
where $f_{\chi^2}(x; \nu)$ is the probability distribution function
of a $\chi^2(\nu)$ random variable  where $\nu=l$ is the number of free parameters
of the nested component,
and $f_{N}(x; 0,\sigma^2)$ is the
probability  distribution
function of an $N(0,\sigma^2)$ random variables, with $\sigma^2=\hat{\sigma}^2_{\Delta C}$
representing the design variance. 
This parent distribution will be referred to as the \emph{$B(\nu,\sigma^2)$ distribution} or the \emph{overdispersed $\chi^2$ distribution}
with parameters $\nu$ and $\sigma^2$.
 Such convolution
 does not in general lead to an analytic expression for the distribution function $f_B$,
 and therefore numerical calculations are required.~\footnote{The convolution
can be readily estimated, for example, using the \texttt{scipy.signal.fftconvolve} routine in \texttt{python},
which performs a convolution using a fast Fourier transform on the discretized distributions.}

\begin{figure}
  \includegraphics[width=3.5in]{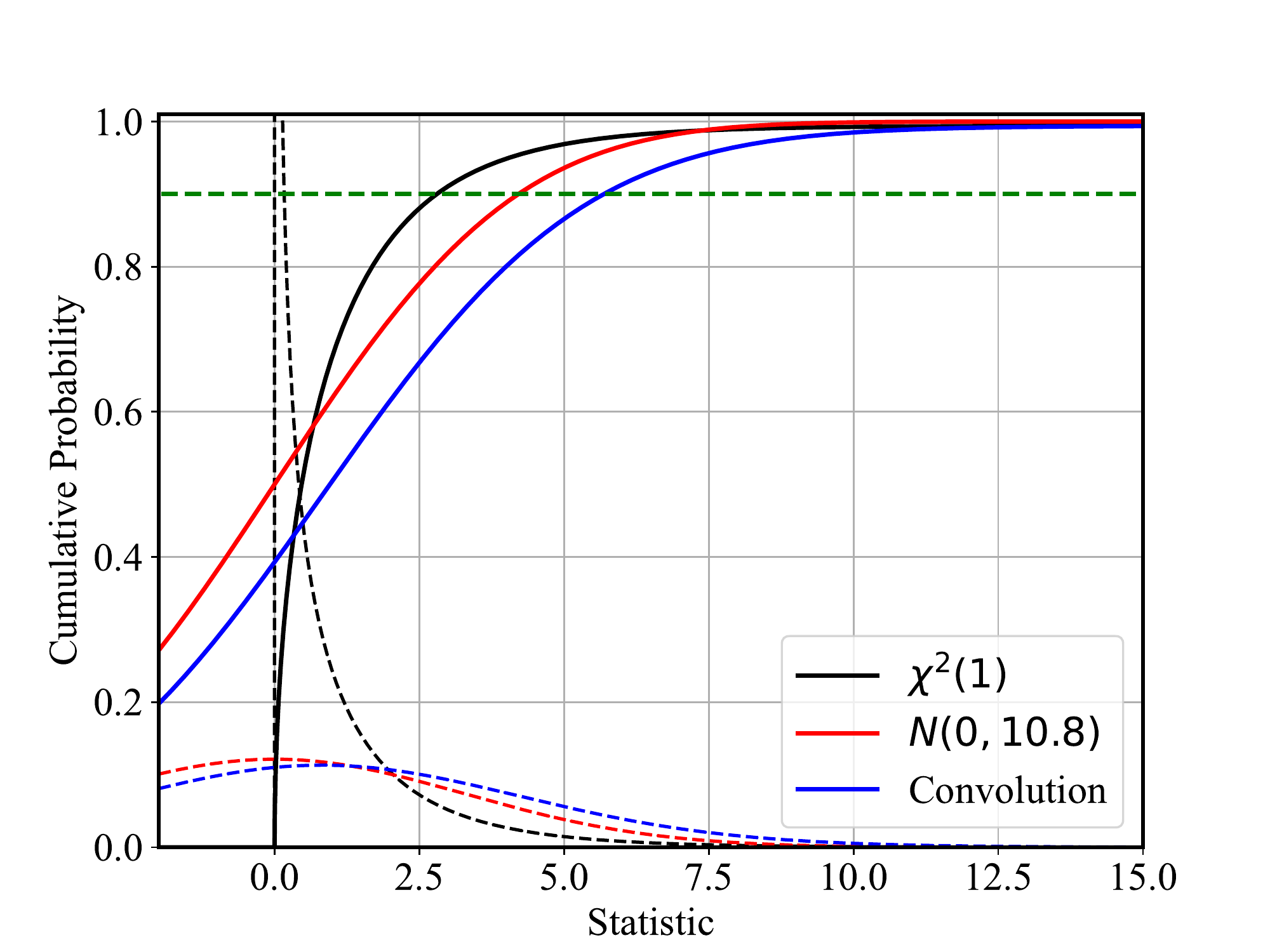}
  \caption{Distribution of the overdispersed $\chi^2$ distribution, resulting from the
  convolution of Equation~\ref{eq:conv}, for $\nu=1$
  degree of freedom of the $\chi^2$ distribution, and a value $\sigma=3.7$ for the standard
  deviation of a zero--mean Gaussian.}
  \label{fig:convolution}
\end{figure}

The $p$--value associated with a measurement of $z=\Delta C$ is given by
\begin{equation}
  F_B(\Delta C; \nu,\sigma^2)=1-p
  \label{eq:pValueDeltaC},
\end{equation}
where $F_B$ is the cumulative distribution of the overdispersed $\chi^2$ distribution.
The probability distribution function $B(\nu,\sigma^2)$ is illustrated in Fig.~\ref{fig:convolution}
for a representative case of the parameters.
Critical values for this overdispersed $\chi^2$ distribution, for selected values of the number
of degrees of freedom $\nu$ and of the intrinsic model variance $\sigma^2$,
are shown in Table~\ref{tab:modChiCrit}, which can also be used to 
calculate $p$--values. More comprehensive tables for the overdispersed
$\chi^2$ distribution can be easily obtained via numerical integration of \eqref{eq:conv}.

This $B(\nu,\sigma^2)$ distribution is proposed as the distribution of choice for 
the $\Delta C$ statistic with $\nu$ degrees of freedom, in the presence of an intrinsic model variance that is modelled
as a zero--mean Gaussian distribution with variance $\sigma^2$, according to \eqref{eq:distrDeltaC}. 
The parameter $\sigma=\hat{\sigma}_{\Delta C}$ represents the intrinsic standard deviation, 
with the meaning that the range $C_{\mathrm{min}} \pm \hat{\sigma}_{\Delta C}$
contains approximately 68\% of the expected intrinsic variability of the $\Delta C$ statistic, as caused by the intrinsic model variance.
This intrinsic `design' standard deviation is calculated according to \eqref{eq:VarDeltaC}, 
where the details of the nested model
component are taken into account. For example, a value of $\hat{\sigma}_{\Delta C}=1$ 
for a $\Delta C$ statistic with $\nu=1$ degree of freedom --- e.g.,
for the test of a nested component with 1 additional free parameter ---  
increases the usual 90\% critical value from 2.7 to 3.1. A larger 
value of  $\hat{\sigma}_{\Delta C}=5$ brings the critical value to 7.7, 
with the meaning that any measurement of the $\Delta C$ statistic 
between 2.7 and 7.7 remain \emph{statistically acceptable} at the 90\% confidence level, in the presence of such systematic error.
It is thus clear that underestimating or neglectic systematic errors can lead to erroneous conclusions in the
hypothesis testing of additional nested components.

The overdispersed $\chi^2$ distribution is also appropriate as a parent distribution
for the \cmin\ statistic itself. In fact, as discussed in Sect.~\ref{sec:cminSys}, the $Z=C_{\mathrm{min}}$ statistic
in the presence of systematic errors 
is the sum of two independent variables, the first of which is $X \sim \chi^2(N-m)$. Therefore, the 
overdispersed $\chi^2$ distribution with parameters $\nu=M-n$ and $\sigma^2=\hat{\sigma}^2_{C}$ 
is the distribution of choice for the \cmin\ statistic in the presence of an intrinsic model
variance $\hat{\sigma}^2_{C}$. When $\nu \gg 1$, as is the case for most regressions, 
this distribution asymptotically converges 
to a normal distribution with expectation $\nu$ and variance $2 \nu + \hat{\sigma}^2_{C}$, further 
assuming the data are in the large--count regime. Therefore, for many applications in this regime
it may be convenient to use the normal approximation for the \cmin\ statistic in 
the presence of systematic errors, as discussed in Sect.~\ref{sec:cminSys}.

It is clear that the convolution in \eqref{eq:conv} leads to a distribution that will feature negative values,
unlike in the case of a $\chi^2$ distribution, which is positive definite. This is especially the case
when $\nu$ is a small number and $\sigma^2$ a large value for the intrinsic variance, when
large intrinsic model fluctuation can lead to a $C$ statistic that becomes in fact lower than
\cmin. In practice, the interest is in hypothesis testing with one--sided confidence intervals for
this overdispersed $\chi^2$ distribution, and therefore the negative tail, even when significant as
in the case of Fig.~\ref{fig:convolution}, is not of direct interest to the data analyst.

\begin{table*}
  \caption{Critical values for the overdispersed $\chi^2(\nu)$ distribution, as a function of several
  representative values of the intrinsic variance
  ($\hat{\sigma}_{C}$), and for selected $p$--values. In parenthesis are the corresponding
  critical values for the standard $\chi^2(\nu)$ distribution.}
  \label{tab:modChiCrit}
  \begin{tabular}{l|llllll|llllll|llllll}
	\hline
	\hline
	$\hat{\sigma}_{C} $ & \multicolumn{6}{c}{$\nu=1$} & \multicolumn{6}{c}{$\nu=2$} &\multicolumn{6}{c}{$\nu=3$}\\
		&  \multicolumn{6}{c}{\hrulefill} &  \multicolumn{6}{c}{\hrulefill}&  \multicolumn{6}{c}{\hrulefill}\\
		& \multicolumn{2}{c}{$p=0.9$}&  \multicolumn{2}{c}{$p=0.99$} & \multicolumn{2}{c}{$p=0.999$} &
		\multicolumn{2}{c}{$p=0.9$}&  \multicolumn{2}{c}{$p=0.99$} & \multicolumn{2}{c}{$p=0.999$} &
		\multicolumn{2}{c}{$p=0.9$}&  \multicolumn{2}{c}{$p=0.99$} & \multicolumn{2}{c}{$p=0.999$} \\
    \hline
    1.0 & 3.1 & (2.7) & 7.0 & (6.6) & 11.8 & (10.8) & 4.9 & (4.6) & 9.5 & (9.2) & 14.1 & (13.8) & 6.5 & (6.3) & 11.6 & (11.3) & 16.5 & (16.3) \\ 
2.0 & 4.1 & (2.7) & 7.9 & (6.6) & 12.6 & (10.8) & 5.6 & (4.6) & 10.2 & (9.2) & 14.9 & (13.8) & 7.1 & (6.3) & 12.3 & (11.3) & 17.2 & (16.3) \\ 
5.0 & 7.7 & (2.7) & 13.4 & (6.6) & 18.5 & (10.8) & 8.9 & (4.6) & 15.0 & (9.2) & 19.9 & (13.8) & 10.1 & (6.3) & 16.6 & (11.3) & 22.0 & (16.3) \\ 
10.0 & 14.0 & (2.7) & 24.7 & (6.6) & 33.4 & (10.8) & 15.1 & (4.6) & 25.8 & (9.2) & 33.7 & (13.8) & 16.2 & (6.3) & 27.1 & (11.3) & 35.3 & (16.3) \\ 
15.0 & 20.3 & (2.7) & 36.3 & (6.6) & 49.2 & (10.8) & 21.4 & (4.6) & 37.2 & (9.2) & 48.8 & (13.8) & 22.5 & (6.3) & 38.4 & (11.3) & 50.2 & (16.3) \\ 
20.0 & 26.7 & (2.7) & 47.9 & (6.6) & 65.1 & (10.8) & 27.8 & (4.6) & 48.8 & (9.2) & 64.0 & (13.8) & 28.8 & (6.3) & 49.9 & (11.3) & 65.4 & (16.3) \\ 

    \hline
    \hline
  \end{tabular}
\end{table*}

\subsection{Domains of applicability for the asymptotic distributions}
\label{sec:domains}
The choice to use asymptotic distributions for \cmin\ in \eqref{eq:cminNorm} and  \eqref{eq:distr}, 
and for $\Delta C$ in  \eqref{eq:DeltaC} and \eqref{eq:distrDeltaC}, is motivated by the goal of
obtaining an analytic representation of the distribution of the
associated \emph{overdispersed} statistics that account for the presence of systematic
errors.  The systematic errors are also modeled, for the same reason,
with a simple normal distribution. 
It is clear that these assumptions are not necessary to implement the methods discussed in this paper.
Specifically, the estimation of the intrinsic variance (see Sect.~\ref{sec:intrinsicVariance2})
can also be made using different distributions for the $X$ and $Y$ variables, representing respectively the \cmin\ statistic
and the systematic error. However, in the absence of a simple analytical form for the distribution,
the convolution needs to be carried out numerically, and therefore the simple result of \eqref{eq:sigmai} is no longer guaranteed
to apply. Similarly, the convolution that defines the overdispersed $\chi^2$ distribution in \eqref{eq:conv} can be carried out
between any two distributions, for example if the analyst decides to forgo the assumption of normality for the systematic error,
as was done for example by \cite{lee2011} and \cite{xu2014} in the context of a different framework to address systematic errors.

\begin{figure}
  \centering
  \includegraphics[width=3in]{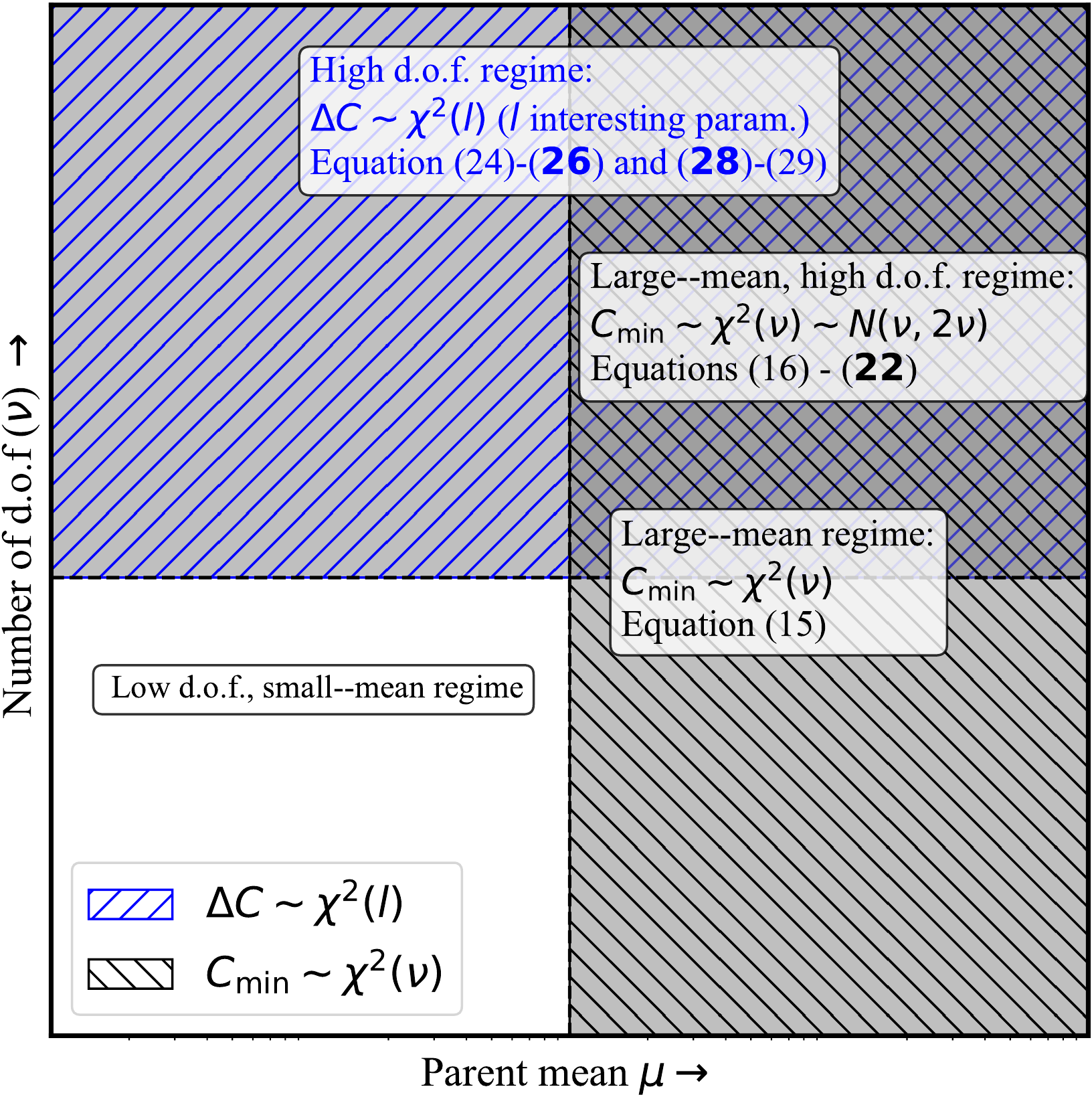}
  \caption{Domains of applicability of key results presented in this paper. Equations corresponding to the
  main results presented in this paper are shown in boldface. 
  The domain of datasets with few datapoints \emph{and} with small parent means
  (lower--left corner) is the portion of parameter space where the results of this paper do not generally apply.}
  \label{fig:box}
\end{figure}
The domains of applicability for the results presented in this section are summarized in Fig.~\ref{fig:box}.
In particular, equation~\eqref{eq:sigmai} applies in the large--mean regime, and for any number of 
degrees of freedom in \cmin, while the overdispersed $\chi^2$ distribution presented in \eqref{eq:conv}
applies for all values of the parent mean, thus even in the low--mean regime, but for a large number
of degrees of freedom. 

The boundaries between the domains are not sharp, and they can be estimated approximately as follows.

(a) In the high degree--of--freedom
regime, the log--likelihood $\Delta C$ statistic is approximately $\chi^2$--distributed, when there is a large number of 
measurements. This result follows from  the \cite{wilks1938} theorem,
which shows that the log--likelihood is $\chi^2$--distributed to within a term of order $1/\sqrt{n}$
(see also Chapter~16 of \citealt{bonamente2022book}). 
As a result, an approximate rule--of--thumb is that $\nu \geq 20$ or so might be sufficient to achieve
a percent--level accuracy, and generally most astronomical datasets
are in this regime.

(b) The large--mean regime where \cmin\ is $\chi^2$--distributed is essentially driven by the 
approximation of a Poisson distribution with a normal distribution of same mean and variance, which is
well satisfied when $\mu \geq 10$ or so. This is shown for example by \cite{kaastra2017} and \cite{bonamente2020}, who 
provide approximations for the mean and variance of the $C$ statistic in the low--mean regime with the aid of numerical
calculations and simulations. 

\section{Application to the X--ray data of \es}
\label{sec:es}

\begin{figure*}
  \centering
  \includegraphics[width=3.2in]{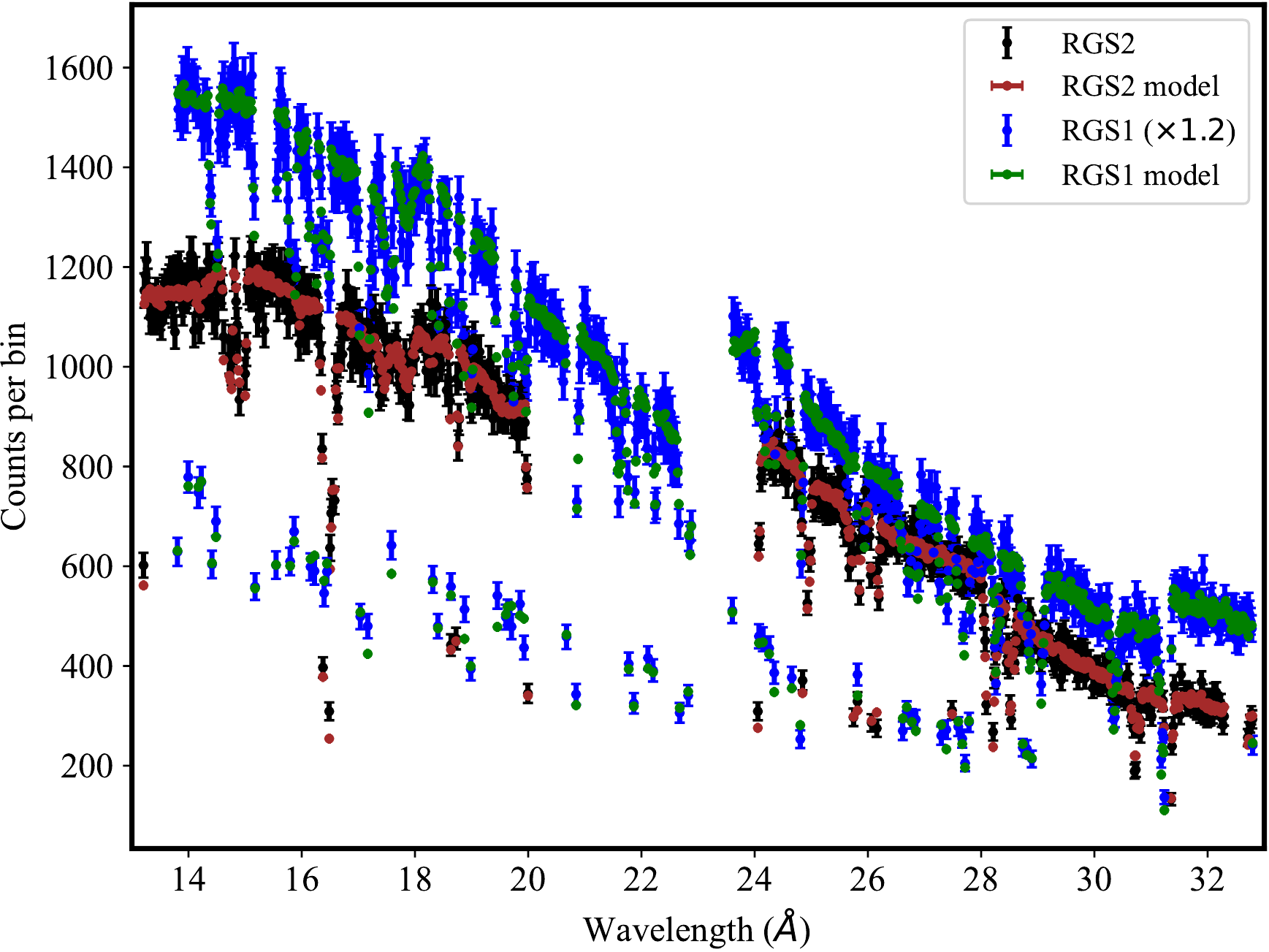}
  \includegraphics[width=3.2in]{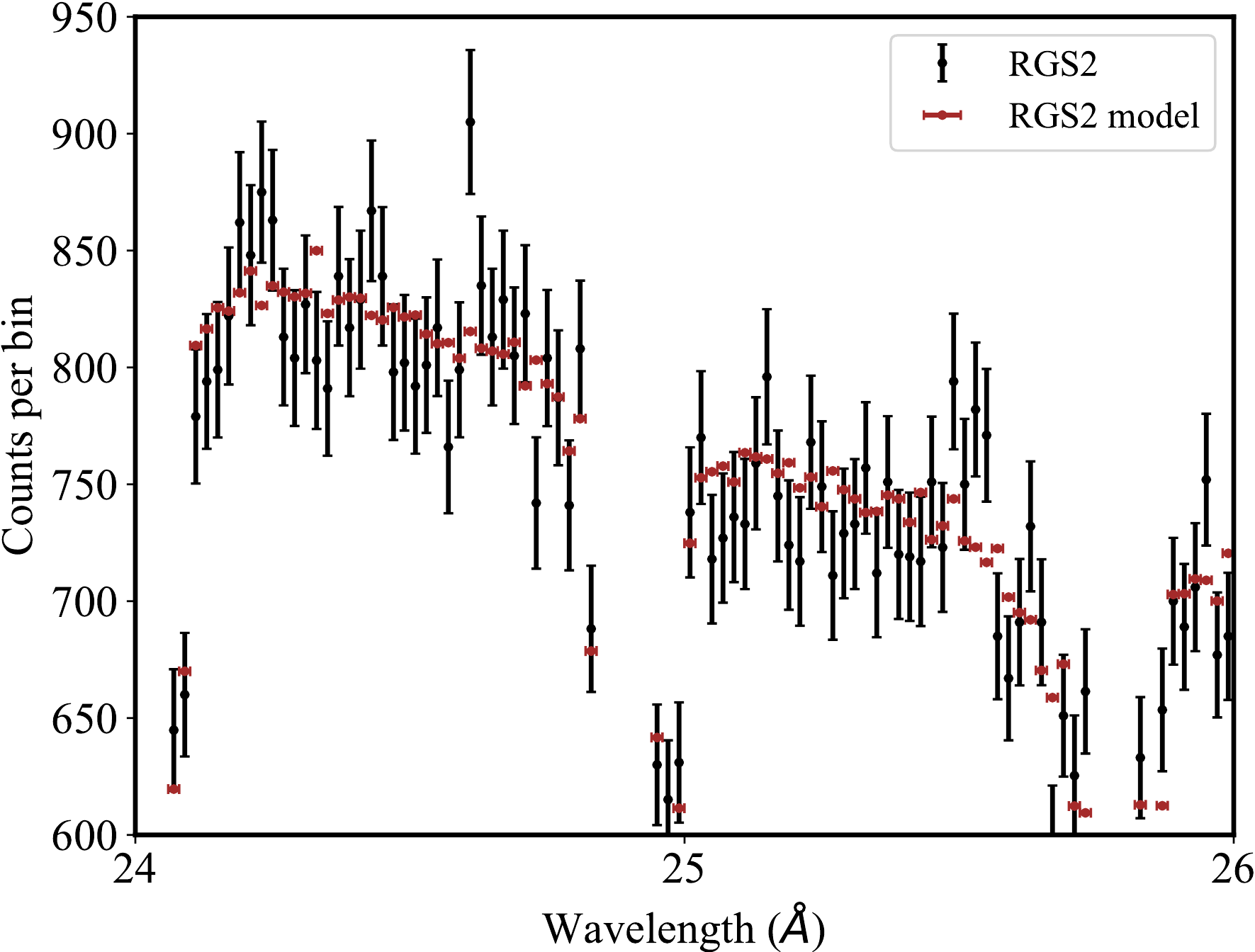}
  \caption{Left: \xmm\ spectrum of the quasar \es, which includes two independent spectra from
  the RGS1 and RGS2 cameras. For clarity, the RGS1 data was shifted by a factor
  of 1.2. Right: Zoom--in of the spectrum in a narrow wavelength interval, to illustrate the data and model.
  For each datapoint, the black dot represents the integer number of counts (source plus background),
  and the error bar is the Poisson error, which is not explicitly used for the \cstat.}
  \label{fig:es}
\end{figure*}

We illustrate the methods described in Sect.~\ref{sec:intrinsicVariance}
to the \xmm\ grating spectra of \es, which are described in a  separate paper \citep{spence2023}. 
These data were used by \cite{nicastro2018}
to detect \ovii\ absorption from the intervening warm--hot intergalactic medium,
and they are representative of the data model discussed in this paper. 
The spectrum is shown in Figure~\ref{fig:es}, where the data points represent 
the total (integer) number of counts $y_i$ in a given bin, and the horizontal markers represents
the best--fit model $\hat{\mu_i}$. Several wavelength ranges where excluded because of
poor calibration, and several bins adjoining excluded ranges have a smaller size than the
default 20~m\AA\ bin size that is used in the data reduction process, thus the lower
number of counts. 

For the purposes of this paper, the details of the continuum modelling are not
important, and
the main statistics of interest to the present analysis are summarized in Table~\ref{tab:fit}.
The methods of hypothesis testing and the determination of the systematic error
only rely on the numbers presented in that table.
For reference, the analysis is briefly summarized in the following. 
The model used to fit the data comprises a spline continuum with several free parameters
in the 13--33~\AA\ wavelength range, and additional components that model the local background.
Specifically, the spline continuum was parameterized in 1~\AA\ 
intervals, to each of which corresponds one free parameter.
For these data, the background flux is $\sim 10$\% of the source's flux
on average over the
interesting wavelength range, and it has much smaller fluctuations that the source because
it is collected from a larger area compared to that of the point source.
The spectral analysis was conducted with the \spex\ 3.0 fitting package \citep{kaastra1996}, which uses
the integral number of counts (source plus background) to calculate the $C$ statistic.
A full description of the data and data analysis is provided in a companion paper
\citep{spence2023}, including an analysis of the background and its effects on
the calculation of the fit statistics. 

\subsection{Hypothesis testing and the need for systematic errors}
\label{sec:hypothesis}
The method of hypothesis testing with the \cstat\ is described in detail
in \cite{kaastra2017} and in \cite{bonamente2022book}.
For this application, the method is simplified considerably by the large--$N$ and large--mean
regime of these data, where the \cmin\ statistic is expected to be distributed like a $\chi^2(N-m)$ distribution,
where $m$ is the number of free parameters in the fit.
Given that a $\chi^2$ random variable with $\nu$ degrees of freedom has
a mean of $\nu$ and a variance of $2 \nu$, the measured value of the statistic
needs to be compared with confidence intervals 
of this distribution. 

For convenience, in the following the range  $\nu \pm \sqrt{2 \nu}$ is used 
to indicate the range within one standard deviation of the mean of a $\chi^2$ distribution with $\nu$ degrees of freedom. 
It is useful to point out that this range for \cmin\
is somewhat different from what is reported by the \spex\ package, which instead assumes that
the expectation of the fit statistic is equal to the sum of the $N$ expectations for each of the
contributing terms of \eqref{eq:cstat}, asymptotically $N \pm  \sqrt{2 N}$ for large means,
with no reference to the number of adjustable parameters \citep{kaastra2017}. 
Although there is only a small difference between the 
two ranges in Table~\ref{tab:fit}, 
the reader is referred
to Sect.~16.3 of \cite{bonamente2022book} and to \cite{bonamente2020}, where the
method of hypothesis testing with the \cstat\ is described in detail, including what is known
concerning the reduction--of--degrees--of--freedom for the \cstat, 
based on a theorem of \cite{wilks1938}. As already remarked earlier in the paper (see Sect.~\ref{sec:Cmin}),
the issue is also addressed in \cite{mccullagh1986}, who
showed that the first--order approximation for the mean of \cmin\ is $N$, while a second--order approximation
provides the $m$ correction, to yield an expectation of $\nu=N-m$. Accordingly, in this analysis we assume
a reduction--of--degrees--of--freedom result for the \cstat, with a parent mean of $\nu$ for the \cmin\
statistic in the large--count limit.

Hypothesis testing is based on
 the $p$ value of the fit, defined as the probability that the
 measured statistic (in this cases \cmin) has a value greater or equal than the
 measured value, under the hypothesis that the data are drawn from the parent model, i.e.,
 the model under consideration is an accurate representation of the data.
For the overall \cmin=1862.7 statistic and its expectation based on the $\chi^2(N-m)$
asymptotic distribution reported in Table~\ref{tab:fit},
the measured value falls at a value of $z=+7.1$ standard deviations from the parent mean
under the null hypothesis of $1478\pm54.4$, which corresponds to
an infinitesimally small $p$--value. Similar conclusions apply to the
two fit statistics for the RGS1 and RGS2 data considered separately. 
It is  therefore clear that the measured fit statistics are not compatible with
the null hypothesis.

This situation of a higher--than--expected fit statistic --- 
be it the \cstat\ or $\chi^2$ --- is 
quite common in X--ray astronomy. 
In cases
where there are extended wavelength regions where the best--fit model is significantly above or below the
data (of which there are numerous examples in the literature, such as Fig.~4 of \citealt{bonamente2003}
for a different dataset),
the best course of action is to deem the current model \emph{unacceptable}, and thus discard it and
try an alternative model. In this case, however,  there are no systematic trends
in the residuals, i.e., the best--fit model \emph{appears} to be  an overall reasonable approximation to the data.
The analyst is thus presented with two alternative choices: (a) to discard this model, or 
(b) to investigate whether there may be additional sources of uncertainty 
that might have contributed to the mis--match between measured and expected value for the
fit statistic. In this section we follow the latter approach,
and use the method of Sect.~\ref{sec:intrinsicVariance} 
to estimate the amount of intrinsic variance in the model
that brings the model to an acceptable degree of agreement with the data.
This method assumes that there is no structure in the residuals, i.e., the additional variance in
the data is somewhat uniformly distributed among all datapoints. When following this approach, the analyst must thus ensure
that there is an analysis of the residuals that supports this assumption. For the data at hand, the analysis 
 provided in \cite{spence2023} indeed indicates that there is no significant structure in the residuals.

It is also useful to remark, at this point, that in the astrophysical literature
it is common to report a \emph{reduced} value for the fit statistic (typically the value
of $\chi^2_{\text{red}}=\chi^2_{\text{min}}/\nu$), and deem the fit acceptable if
such reduced value is 'close' to one. A statistically more sound and quantitative
approach is to report the $p$ value of the fit, as explained earlier in this section,
since the proximity of a reduced value for $\chi^2$ or for the
$C$ statistic is a strong function of the number of datapoints or of degrees of freedom.
For the overall \cmin\ statistic and its expectation based on the $\chi^2(N-m)$
asymptotic distribution reported in Table~\ref{tab:fit}, the $p$ value was 
found to be essentially zero,
although the value of the reduce \cmin\ statistic is 1.29, and it may be erroneously deemed
`close' to unity.

Prior to the estimation of systematic errors according to the methods of
Sect.~\ref{sec:intrinsicVariance}, we use
Equations~\ref{eq:VarNB} and \ref{eq:phiHat} to estimate the overdispersion parameters in the data.
For the NB1 parameterization, we estimate
\begin{equation}
  \hat{\phi}=
  \begin{cases}
    1.58\pm0.05\; \text{ for RGS1 data}\\
    1.35\pm0.05\; \text{ for RGS2 data}
  \end{cases}
  \label{eq:phiHatResults}
\end{equation}
where the uncertainties are based on a simple error--propagation or delta method, assuming that the
estimated mean values are known accurately. Given that $\hat{\phi}>1$, the estimate confirms
the presence of an additional source of variance in the data.
For the NB2 parameterization, we estimate
\begin{equation}
  \hat{\alpha}=
  \begin{cases}
    1.5\pm0.3 \times10^{-3} \; \text{ for RGS1 data}\\
    0.6\pm0.2 \times10^{-3} \; \text{ for RGS2 data}
  \end{cases}
  \label{eq:alphaHatResults}
\end{equation}
where the $\alpha$ parameter multiplies the square of the mean number of counts per bin,
thus the lower absolute value compared with the NB1 model. 
As for the previous parameterization, the value $\hat{\alpha}>0$
indicated overdispersion.
The $\hat{\phi}$ and $\hat{\alpha}$ parameters provide quantitative evidence for overdispersion in the data,
relative to the best--fit model obtained with the usual Poisson log--likelihood. The parameters, however,
do not address the effect of overdisperion or intrinsic model variance 
on the fit statistic, which is studied in the following
using the methods presented in Sect.~\ref{sec:intrinsicVariance}.

\subsection{Estimate of systematic model uncertainty with the \cstat}

\label{sec:sys}

\begin{table}
  \centering
  \caption{Fit statistics for the \xmm\ spectrum of \es}
  \label{tab:fit}
  \begin{tabular}{lll}
    \hline
    \hline
    Statistic & Value & Notes \\
    \hline
    \multicolumn{3}{c}{Combined RGS1 and RGS2 data}\\
	\cmin & 1862.7 &  ($1861.8$ from \protect\citealt{spence2023})~$^\star$\\
	$\chi^2$ & 1730.6 & (not used in the minimization) \\ 
	$N$ & 1526 & Number of data points\\
	$m$ & 48   & Number of free parameters\\
	$f=N-m$ & 1478 & Number of degrees of freedom \\
	Expected \cstat & $1526.4  \pm$ 55.3 & As reported by \spex \\
			& $1478\pm54.4$ & According to the $\chi^2(N-m)$ approx. \\
	\hline
	\multicolumn{3}{c}{RGS1 data}\\
	\cmin & 1023.5 & \\
	$N_1$  & 801 & \\
	$\chi^2$ & 1023.5 \\
	$m_1$ & 48 & Resulting in $N_1-m_1=753$ d.o.f.\\
	Expected \cstat & $753.0 \pm 38.8$ & According to the $\chi^2(N-m)$ approx. \\
	\hline
	\multicolumn{3}{c}{RGS2 data}\\
	\cmin & 788.2 & \\
	$N_2$ & 725 & \\
	$\chi^2$ & 788.2 \\
	$m_2$ & 44 & Data in 20--24 \AA\ range ignored, \\
		&	& resulting in $N_2-m_2=681$ d.o.f. \\
	Expected \cstat & $681.0\pm36.9$& According to the $\chi^2(N-m)$ approx. \\
\hline
\hline
  \end{tabular}
  \flushleft {\footnotesize  $\star$ The small difference
  is due to approximations made by the \texttt{SPEX} software, and they are irrelevant for our analysis.}

\end{table}

\subsubsection{Preliminary considerations}
 Given the values of the statistics reported in Table~\ref{tab:fit},
it is reasonable to proceed independently with an analysis of the two \xmm\ instruments. In so doing, 
this application  also illustrates the determination of systematic errors from more than
one independent dataset. First of all we point out that the model in use
does not allow for a free normalization between the models applied to RGS1 and RGS2. 
We therefore begin the investigation for the origin of the poor fit statistic by repeating the same 
fit only to RGS1 and RGS2 data. From this exercise,
we obtained statistically equivalent fits as the ones reported at 
the bottom of Table~\ref{tab:fit}, which refers to the contributions to the statistic from
the usual joint RGS1 plus RGS2 regression of \cite{spence2023}. We  thus conclude that 
a cross--normalization error between the two instruments is not the origin for the poor \cmin\ value.
We therefore proceed with the numbers in Table~\ref{tab:fit} as the basis for our estimate of the 
systematic error.

\subsubsection{Determining the design value for the variance of \cmin}

The starting point is to determine a `design' value for the intrinsic variance of the \cmin\ statistic
required to bring agreement with the measured value. Qualitatively speaking, such value must bring
agreement between the distribution function of the \cmin\ statistic and the measured value, currently at odds (e.g.,
RGS1 has a measured value of 1023.5 versus an expectation and standard deviation
of $753.0 \pm 38.8$). 
With $F_C$ the cumulative distribution function of the parent distribution of the 
\cmin\ statistic, one may require that
the measured value of the statistic satisfies
\begin{equation}
  F_C(C_{\text{min}})=1-p,
  \label{eq:FC}
\end{equation}
where $p$ is a high probablity, e.g., $p=0.9$ or 0.99. Eq.~\ref{eq:FC} means that the measured value \cmin\ is
required to be the $p$ quantile, with a residual probability of just $1-p$ to exceed the measured
value, according to the sampling distribution of the fit statistic. Equation~\ref{eq:FC} uses
a
single--sided rejection region of $C_{\text{min}}>C_{\text{crit}}$, as is customary for hypothesis testing with
the $\chi^2$ distribution. In \eqref{eq:FC}, $C_{\text{min}}$ is the actual measurement of the statistic with the data
at hand.
Given that the problem of intrinsic variance arises when the statistic is sufficiently large and in excess
of its expectation,  \eqref{eq:FC} is understood as featuring a measurement
of the statistic at the boundary of the rejection region, i.e., $C_{\text{min}}=C_{\text{crit}}$.

When the \cmin\ statistic
is normally distributed (which is the case for a large number of measurements and in the large--count limit, 
as is the case for this application), then \eqref{eq:FC} is equivalent to the familiar requirement that the measured 
value exceeds the mean by a predetermined number of standard deviations,
\begin{equation}
  C_{\text{min}} = \text{E}[C_{\text{min}}]+ \beta \times \sqrt{\text{Var}(C_{\text{min}})}
  \label{eq:FC2}
\end{equation}
with $\beta=1.28$ or 2.33 for values of $p=0.9$ or 0.99. 
Equations~\ref{eq:FC} or \ref{eq:FC2}
must be solved for the value of the intrinsic variancs $\hat{\sigma}^2_C$, with
\begin{equation}
  \text{Var}(C_{\text{min}})=\sigma^2_C +\hat{\sigma}^2_C
  \label{eq:Var}
\end{equation}
and $\sigma^2_C$ indicating the parent variance of the statistic without accounting
for the intrinsic model uncertainty. 
The design value for the intrinsic variance is immediately calculated from \eqref{eq:FC} (or \eqref{eq:FC2} assuming
Gaussian distribution for the fit statistic) and \eqref{eq:Var} as
\begin{equation}
  \hat{\sigma}^2_C = \dfrac{\left(C_{\text{min}} - \text{E}[C_{\text{min}}]\right)^2}{\beta^2} - \sigma^2_C.
  \label{eq:intVar}
\end{equation}

For the values of the statistics in Table~\ref{tab:fit} and with a choice
of $\beta=2.33$ (corresponding to a 99\% one--sided confidence level), 
Eq.~\ref{eq:intVar} yields the following
values of the intrinsic variance for the RGS1 and RGS2 data separately:
\begin{equation}
  \begin{cases}
    \hat{\sigma}_{C,1} =&109.4,\; \text{ for a prediction of $753 \pm 116$;} \\
	\hat{\sigma}_{C,2} =&57.0,\; \text{ for a prediction of $681 \pm 67.9$.} \\
    \end{cases}
    \label{eq:sigmaC1}
\end{equation}
When these intrinsic variances
are combined for the entire RGS1+RGS2 spectrum, they result in
\begin{equation}
\hat{\sigma}_{C}=123.4,\; \text{ for an overall prediction of $1478 \pm 134.8$.}
  \label{eq:sigmaC2}
\end{equation}

Equations~\ref{eq:sigmaC1} and \ref{eq:sigmaC2} are the sought--after estimates for the 
`design variances'
required for statistical agreement between the best--fit model
and the data, according to the maximum--likelihood Poisson \cmin\ goodness of fit statistic.
These design variances were added to the original `statistical' variances of \cmin\ (i.e., the
variances based on the Poisson counting statistics of the data) according the \eqref{eq:Var}, to yield
the overall variance of each statistic.
Notice how the $z$-score of the overall \cmin\ statistic is now $z=2.85$, 
given the usual measurement of \cmin=1862.7 and 
the revised expectation of $1478 \pm 134.8$,
corresponding to 
a one--sided null hypothesis probability to exceed the measured value of $p\simeq0.002$.

\subsubsection{The estimate of the intrinsic variance}
\label{sec:intVarResult}
With these design values for the intrinsic variance of the fit statistic, the use of \eqref{eq:sigmai}
provides the values of the intrinsic model error. 
The average number of counts is respectively 728.6 for RGS1, 756.2 for RGS2, and 741.7 for the
combined dataset. Accordingly, the method yields
\begin{equation}
  \dfrac{\sigma_{\text{int},i}}{\hat{\mu_i}} = \begin{cases} 0.072\, \text{ for RGS1}\\
     0.039\, \text{ for RGS2}\\
     0.058\, \text{ for combined RGS1 and RGS2}.
  \end{cases}
  \label{eq:results}
\end{equation}
The interpretation of these numbers is that there is a level
of systematic uncertainty in the best--fit models in the amount of
respectively 7.2\%, 3.9\% and 5.8\% in RGS1, RGS2 and for the combined data.

This method has therefore provided estimates
of the intrinsic model variance based on simple analytical methods, for the
individual RGS1 and RGS2 spectra and for the combined spectrum.  These variances are estimated as fractional errors
of the best--fit model in each bin according to \eqref{eq:results}, assuming that
the relative systematic errors are uniform across the spectrum.
 It is necessary to compare these numbers with known systematic errors of the
\xmm\ instrument. \cite{spence2023} shows that the typical systematic errors are of order of a few percent
of the number of counts in a bin of the size used for this analysis \citep[see also][]{marshall2021}. 
The results of the analysis
presented in this paper, with 4-7\% fluctuations relative to the mean,
is therefore consistent with the known level of systematic errors in the instrument. 

Finally, it is useful to compare the values of 
estimated overdispersion parameter $\hat{\phi}$ from \eqref{eq:phiHatResults} with the intrinsic model variance 
in \eqref{eq:results}.
The overdisperion parameter
$\hat{\phi}=1.3-1.6$ 
models the additional variance in each measurement, with a value of $\phi=1.0$ representing a
variance equal to the mean. The square root of the variance $\omega_i=\hat{\phi} \,\mu_i$
(or
$\omega_i=\mu_i+\alpha \mu_i^2$ for the alternative parameterization)
needs to be divided by the mean number of counts in the bin ($\mu_i$) in order to
assess the additional level of fractional systematic errors in the data implied by the extra variance.
For a typical count rate in these data of several hundred counts per bin, the
estimated values of $\hat{\phi}$ and $\hat{\alpha}$
yield fractional fluctuations that are a few percentage points larger than the case of $\phi=1$, in general
accord with the percent--level intrinsic model variance estimated in \eqref{eq:results}.


\subsection{Systematic errors in the data with the $\chi^2$ statistic and other considerations}
Given that the analysis of Sect.~\ref{sec:sys} indicates an intrinsic model uncertainty of
order $\sim 5$\%,
it is natural to ask the question of how such uncertainty, if applied to the \emph{data} instead,
would affect the $\chi^2$ statistic. The reason to pursue this comparison is that of providing
a consistency check with a more traditional, albeit less accurate, method of
assessing systematic errors in the data in the presence of an unacceptable fit statistic.

To this end, we apply the average intrinsic errors  of \eqref{eq:results} to each data point,
and add them in quadrature to the usual Poisson errors. With this change in the data errors, the 
$\chi^2$ statistics are modified to
\[
  \chi^2 = \begin{cases}
    874.8\, \text{ for RGS1 } (\Delta \chi^2=-68.5)\\
    626.1\, \text{ for RGS2 } (\Delta \chi^2=-101.82)\\
    1500.9\, \text{for RGS1+RGS2 } (\Delta \chi^2=-170.4).
  \end{cases}
\]
with respect to the values of Table~\ref{tab:fit}. The main result is that a systematic
error added to the data in the amount of 6.5\% results in a decrease of $\Delta \chi^2=-170.4$,
bringing the statistic in closer agreement with its expectation, for the given number of
degrees of freedom ($\nu=1478$). This exercise therefore indicates that the values of
the intrinsic model variance estimated from the Poisson data and the \cstat\ 
with the novel method described in Sec.~\ref{sec:sys} are {reasonable}, in that
a systematic error of the same magnitude applied to the data and to the $\chi^2$ statistics
bring this statistic closer to its expectation, in a manner similar to the change in the \cmin\
statistic using the intrinsic model variance.

\begin{figure}
  \includegraphics[width=3in]{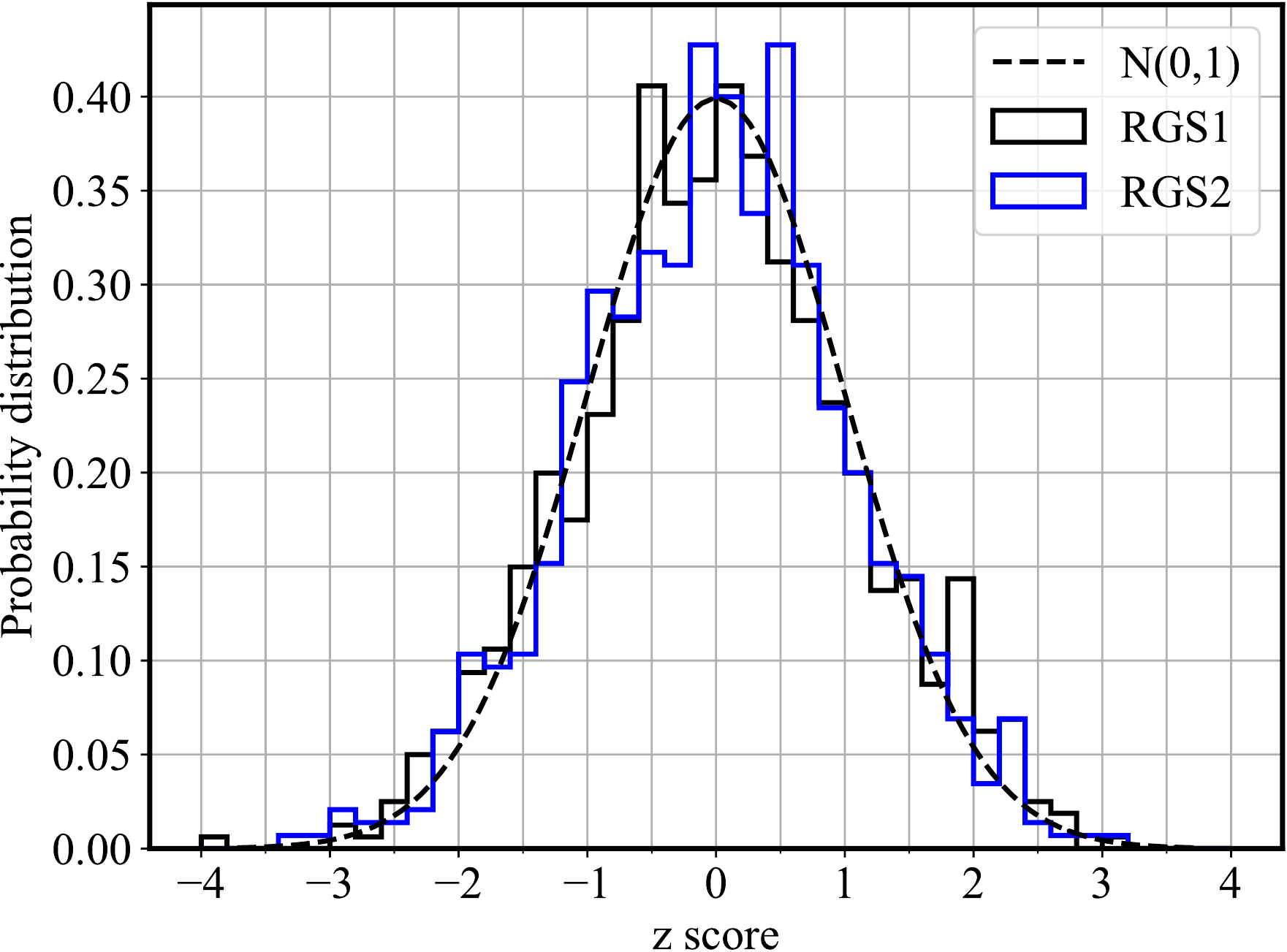}
\caption{Distribution of $z$--scores for the RGS1 and RGS2 data. The mean and standard deviation of the two samples
  are respectively 0.00 and 1.08 for the RGS1 z--scores, and -0.01 and 1.04 for the RGS2 z--scores.}
  \label{fig:z}
\end{figure}
It is also instructive to plot the probability distribution of the $z$--scores associated with 
the residuals from the best fit, using the usual Poisson counts as the data variance.
These distributions, shown in Fig.~\ref{fig:z}, do not show any significant departure from
the standard normal distribution, which is the expected distribution under the null
hypothesis. This observation is also confirmed by a Kolmogorov--Smirnov test \citep{kolmogorov1933}, which 
is an empirical distribution function (EDF) test that uses
unbinned data. The one--sample Kolmogorov--Smirnov test statistics are $D_{N}=0.033$ and 0.018 respectively for
the RGS1 and RGS2 data, with corresponding $p$--values  of respectively 0.33 and 0.97, indicating an excellent agreement
between the data and the parent model.~\footnote{The test is described, e.g., in Ch.~19 of 
\cite{bonamente2022book}, and it is implemented in the \texttt{kstest} python script freely available
via the \texttt{scipy.stats} package.} 
The Kolmogorov--Smirnov test is therefore not sensitive for the detection of systematic errors
at the level of a few percent, i.e., at the level present in the data of this case study.
The Anderson--Darling EDF statistic \citep{anderson1952, stephens1974} for the test of normality of the distributions has 
values of $A^2=0.32$ and 0.21 respectively for
the RGS1 and RGS2 data, with a corresponding 99\% critical value of 1.09 for both data, indicating that the null
hypothesis of normally distributed data cannot be rejected. Therefore, similar to the one--sample Kolmogorov--Smirnov test,
the Anderson--Darling test cannot find deviations from normality in the z--scores.~\footnote{The Anderson--Darling test
is implemented as \texttt{anderson} in the \texttt{scipy.stats} package.}

\subsection{Effect of intrinsic errors on
a nested component with the $\Delta C$ statistic}
\label{sec:1ESDeltaC}

Equation~\ref{eq:VarDeltaC} provides the means to estimate the uncertainty in the $\Delta C$ statistic
due to the intrinsic model variance. 
For the case study with the \es\ \xmm\ data, the relevant statistics are reported in Table~\ref{tab:DeltaC},
reproduced from \cite{spence2023}. The regressions  used for the detection of a 
nested line component are now performed in narrow bands of $\pm 1$~\AA\
around the expected line features, instead of a broad--band fit (as for the results of Table~\ref{tab:fit}).
For the data of Table~\ref{tab:DeltaC}, the baseline model is composed by a simple power--law model, supplemented
by a Gaussian line model (\texttt{line} in SPEX). The line models provide a deviation from the smooth continuum
at the level of a few percent, and therefore it is useful to determine how an intrinsic error at a similar
level affects the significance of detection of these features. An illustration of the data
leading to the reults of Table~\ref{tab:DeltaC} is provided in Figure~4 of \cite{spence2023}.

\begin{table*}
  \caption{$\Delta C$ statistics from the fit to a power--law model plus a Gaussian line feature for the \xmm\ data of \es,
  reproduced from \protect\cite{spence2023}.  The `$p$--value' columns correspond to
  the usual $p$ values of the regression, and the '$P$ value' columns correspond to the $p$--value with the inclusion
  of uncertainties associated with the blind search of lines at an unknown redshift.}
  \label{tab:DeltaC}
  \begin{tabular}{l|llll|ll}
    \hline
    \hline
    Target line & $\Delta C$ & $\nu$ & \multicolumn{2}{c} {$p$--value} & \multicolumn{2}{c}{$P$  value with redshift trials} \\
    \multicolumn{3}{c}{\hrulefill} &  \multicolumn{2}{c}{\hrulefill} &  \multicolumn{2}{c}{\hrulefill}\\
	&	&	&(no sys. error) & overdispersed  $\chi^2$ distr. & (no sys. error) & overdispersed $\chi^2$ distr.  \\
    \hline
    \ovii\ $z=0.1876$ & 6.6 & 1 & 0.010 & 0.066 & \nodata & \nodata\\
    \ovii\ $z=0.4339$ & 29.9 & 2 & $3.2 \times 10^{-7}$ & $2.6 \times 10^{-5}$ & $4.1 \times 10^{-5}$ & $3.3\times 10^{-3}$ \\
    \hline
    \ovii\ $z=0.3551$ & 8.2 & 2 & 0.017 & 0.880 &  \multicolumn{2}{c}{\nodata} \\
    \hline
    \hline
  \end{tabular}
\end{table*}

The largest uncertainty in the use of \eqref{eq:VarDeltaC}
is the number $n$ of independent datapoints in the sum. In the case of an absorption line--like model
such as the Gaussian profile of \texttt{line}, a simple method to determine the number of independent data
points is by calculating the equivalent width of the absorption--line feature. In all cases, the 
equivalent width is smaller than the instrument's resolution, and therefore a simple estimate is $n=1$, meaning that
the model affects just one datapoint, also
leading to a conservative (i.e., the smallest possible) estimate of the intrinsic variance. With a
characteristic number of counts of respectively 800, 500 and 400 per bin for the three lines,
and an intrinsic model error of $\sigma_{\text{int,i}}/\hat{\mu}_i = 0.058$ as estimated 
in Sect.~\ref{sec:intVarResult}, 
\eqref{eq:VarDeltaC} leads to an estimate of

\begin{equation}
  \hat{\sigma}^2_{\Delta C} =
  \begin{cases}
    \begin{aligned}
      10.8\; &\text{ for line 1}\\
      6.7\; &\text{ for line 2}\\
      5.4\; &\text{ for line 3}.\\
    \end{aligned}
  \end{cases}
\end{equation}

Following the modified $\chi^2$ distribution for the $\Delta C$ statistic in the presence
of intrinsic errors described in Sect.~\ref{sec:DeltaC}, it is now possible to
determine whether the three lines in Table~\ref{tab:DeltaC} remain statistically significant,
after accounting for the effect of systematic errors. The results are reported in the `$p$--value'  columns 
of Table~\ref{tab:DeltaC}. 

The statistical significance of the first line is
reduced from a $p$--value of 1\% to $\sim 7\%$. It is therefore possible to conclude that, \emph{if} the
additional variance of the data relative to the best--fit models of Table~\ref{tab:fit} is interpreted 
as an additional source of systematic error, as the methodology presented in this paper posits, 
then the absorption line at $z=0.1876$ is likely caused by 
a statistical fluctuation, and unlikely to be a genuine celestial signal. In this paper, it is not
appropriate to further comment on the astrophysical significance of this putative absorption line, since
the main aim of this report is to establish a method of analysis for systematic errors. The conclusion is thus
that sources of systematic errors in Poisson data such as the spectra analyzed in 
\cite{spence2023}
can be easily addressed, and that their impact can be significant.

The second line, originally reported by \cite{nicastro2018}, is also reduced in significance 
by  factor of nearly $\sim 100$.
For this absorption line, it is necessary to consider the fact that its redshift was identified serendipitously, thus
the additional degree of freedom in the analysis (i.e., the center wavelength of the absorption line).
For such serendipitous sources, \cite{kaastra2006} and \cite{bonamente2019} describe a method
to address the several independent opportunities to detect a fluctuations, known as \emph{redshift trials}.
Given the large number of independent opportunities to detect a random fluctuations, 
estimated in \cite{spence2023} to be of order $N \simeq 100$,
the method
described in \cite{bonamente2020} results in corrected $p$--values that are reported in the right--most
columns of Table~\ref{tab:fit}~\footnote{For a small value of $p$, the method is
approxiamtely equal to $P \simeq p\,N$.}.
 Using the revised $p$--value from the overdispersed $\chi^2$ distribution,
the redshift trials--corrected $P$ value is now approximately 0.003, which corresponds approximately to 
a $\pm 3 \sigma$ two--sided confidence level for a normal distribution. It is therefore clear that 
even a seemingly `strong' detection of a nested model component ($\Delta C = 29.9$ for 2 degrees of freedom)
can in fact become far less significant due to the effect of percent--level systematic errors.

The same analysis is not warranted
for the third line, given that it is not statistically significant even without accounting for the intrinsic variance
(probability of 88\% according to the overdispersed $\chi^2$ distribution).
The redshift trials correction does not apply to the first line, given that its central wavelength was
fixed at its known value, and therefore it features no redshift trials.


\subsection{Summary of steps for a typical implementation}
\label{sec:algorithm}
In this section we provide a short summary of the steps for a typical  implementation of the methods presented in
this paper, with the goal of estimating the systematic errors in Poisson count data from the
maximum--regression of a parametric model with the \cmin\ statistic, 
and for the assessment of the significance of a nested
model component with the $\Delta C$ statistic.

1. Determination of the design variance $\hat{\sigma}^2_{C}$ from \eqref{eq:intVar}, where \cmin\ is the measured 
statistic from the parametric regression, with $\nu$ degrees of freedom. This step requires the choice of
a value of $\beta$ for the agreement between the measured statistic and its parent distribution. 
The expectation of the \cmin\ statistic is $\nu$
in the large count regime, and a discussion for the low--count regime was provided in Sect.~\ref{sec:hypothesis}. 

2. The fractional systematic error $\sigma_{\text{int,i}}/\hat{\mu_i}$ is estimated from \eqref{eq:sigmai}, where $N$ is the number of independent
Poisson datapoints, and $\hat{\mu_i}$ is the best--fit model for the i--th datapoint. 
This number represents the fractional uncertainty associated to the best--fit model of each datapoint.

3. When the significance of a nested component  needs to be assessed, 
the first step is the determination or estimation of the number $n$ of bins where the nested model is present,
and then the variance $\hat{\sigma}^2_{\Delta C}$ estimated according to \eqref{eq:VarDeltaC}.

4. The $\Delta C$ associated to the additional nested component with $\nu=l$ additional parameters  is then compared to the parent overdispered
$\chi^2$ distribution $B(l,\hat{\sigma}^2_{\Delta C})$ defined in \eqref{eq:conv}. Typically this is achieved via
numerical solution of the equation \eqref{eq:pValueDeltaC} which gives the $p$--value associated with the $\Delta C$ statistic.

All these steps require minimal computational effort, given that most results  are elementary analytical formulas, and the
numerical integration of the overdispersed $\chi^2$ distribution is an elementary task, given its simple unimodal
behavior.

\section{Discussion and conclusions}
\label{sec:conclusions}

This paper has presented a simple quantitative method to estimate 
systematic uncertainties in the maximum--likelihood fit of
Poisson count data to a parametric model, using the \emph{Cash} statistic. This situation is of common
occurrence in the analysis of X--ray spectra from astronomical sources 
of the type presented in the case study of \es\ \citep{spence2023}. This sources is
a bright quasar  where the goodness of fit statistic across an extended wavelength range
is formally unacceptable, yet the data do not have significant and systematic deviations from the
best--fit model.
In such cases, the data analyst 
is faced with the choice to either reject the model altogether, or
to determine the presence of additional source of error that are not
present in the Poisson--distributed data. When the analyst determines that it
is reasonable to pursue the latter avenue, the method developed in Sect.~\ref{sec:intrinsicVariance}
leading to the main result presented in Eq.~\ref{eq:sigmai} provides a simple
quantitative method to estimate an \emph{intrinsic variance} present in the model
that renders the data and best--fit model consistent at a predetermined $p$--value. In the case of the \es\ data presented in this
paper and in \cite{spence2023}, we find that the higher--than--expected \cstat\ for the fit
to a complex parametric model can be naturally explained with the presence of a $\sim 5$\%
systematic uncertainty in the best--fit model. 
The method does away entirely with the more commonly used $\chi^2$ statistic,
which is not appropriate for the regression of integer--count and Poisson distributed data, such
as the majority of astronomical X--ray spectra \citep[as shown in, e.g.,][]{humphrey2009,bonamente2020}.

A key application of the intrinsic model variance is through the effect it has on
the $\Delta C$ statistic, which is a likelihood--ratio statistic that is commonly used to test for the significance of
an additional nested component \citep[e.g.][]{protassov2002}. 
Following the methods described in Sect.~\ref{sec:DeltaC}, it was shown that the intrinsic
model variance leads to an additional contribution to the $\Delta C$ statistic 
that can be modelled as a normal distribution with zero mean and a design variance $\hat{\sigma}^2_{\Delta C}$.
This additional distribution 
 is to be added to the usual
$\chi^2(\nu)$ distribution, which represents the Poisson variability in the $\Delta C$ statistic 
from the nested model component with $\nu=l$ free parameters.
This leads to the introduction of the 
\emph{overdispersed $\chi^2$ distribution} that replaces the usual $\chi^2(\nu)$ distribution
for the test of the additional nested component.
This newly introduced distribution, which is the convolution of the two contributing probability distributions
$\chi^2(\nu)$ and $N(0,\hat{\sigma}^2_{\Delta C})$,
is the parent distribution for the usual $\Delta C$ statistic in the presence of systematic errors,
under the null hypothesis that the nested component is not present, thus leading to a simple
method of hypothesis testing for the nested component. 
With the choice of attributing the additional variability caused by systematic effects
to the model, as opposed to the data, the calculation of the $\Delta C$ statistic itself
remains unchanged, and 
the hypothesis testing for the nested component thus only requires the calculation of critical or $p$--values for this new
overdispersed $\chi^2$ distribution, e.g. of the type reported in Table~\ref{tab:modChiCrit}.

In the case study of the \es\ source,
the $\Delta C$ statistic was used to test for the presence of additional absorption line components.
The $\sim 5\%$ systematic error in the best--fit model results in a
an additional variance in the $\Delta C$ distribution, and in a lower level of significance for
the detection of the lines, as reported in Table~\ref{tab:DeltaC}. For example, the first line under consideration,
which was identified in \cite{spence2023} as a possible \ovii\ resonance line at $z=0.1876$, would have its
significance reduced from a 99\% level to a 93\% level (i.e., $p$--value increasing from 0.01 to 0.07).
The other absorption line that was reported by \cite{spence2023}, and which was previously discovered
by \cite{nicastro2018}, has its significance of detection also greatly reduced by the effect of the intrinsic model variance
(see second line in Table~\ref{tab:DeltaC}), yet it remains above the 99\% confidence level ($p$--value of 0.003).
The use of systematic errors in likelihood--ratio statistics (such as $\Delta C$ or $\Delta \chi^2$)
for the detection of additional nested components has therefore the potential to affect the astrophysical
inference from count data,  such as from high--energy spectra of the type investigated
in this case study. This is especially important in fields
such as the detection of X--ray absorption lines from the WHIM, where the combination of a faint
astrophysical signal and limited resolution in the intruments have combined to produce a number of
detections \citep[e.g.][]{ahoranta2021, ahoranta2020, nevalainen2019, kovacs2019,bonamente2016, ren2014,fang2010,
nicastro2005,fang2007, fang2002} that have the potential to be affected by such systematic errors.

The \emph{Cash} statistic was developed in response to the advent of 
new integer--count astronomical X--ray data,
as the maximum--likelihood fit statistic of choice for Poisson data \citep[e.g.,][]{cash1976,cash1979}.
This statistic is often referred to as the Poisson log--likelihood deviance in other
fields of statistics \citep[e.g.][]{cameron2013}.
Recently, the author also developed a new semi--analytical
method to obtain the best--fit parameters for the linear regression of count data,
a problem that was previously only solved numerically \citep{bonamente2022}. 
The typical implementation of the methods described in this
paper are outlined in Sect.~\ref{sec:algorithm}, and it is expected to be straightforward for most applications.
Given that the \cstat\ is usually the appropriate fit statistic for the maximum--likelihood
analysis of these data, the
method presented in this paper therefore provides an additional tool for the
X--ray astronomer (or any other count--data analyst) who wishes to use the  \emph{Cash}
statistic for their data analysis.

\section*{Data Availability}
The \xmm\ X--ray data associated with the \es\ source are available
via 
\texttt{heasarc.gsfc.nasa.gov/cgi-bin/W3Browse}. Processed data such as spectra
can be obtained from the author upon request. 
Numerical routines in \texttt{python} to implement and reproduce 
the key results of this paper, namely the
overdispersed $\chi^2$ distribution \eqref{eq:conv} and the critical 
values of the distribution in Table~\ref{tab:modChiCrit},  are available
at \texttt{https://github.com/bonamem/overdispersedChi2.}
\bibliographystyle{mn2e}


\begin{thebibliography}{59}
\expandafter\ifx\csname natexlab\endcsname\relax\def\natexlab#1{#1}\fi

\bibitem[{{Ahoranta} {et~al}\mbox{.}(2021){Ahoranta}, {Finoguenov},
  {Bonamente}, {Tilton}, {Wijers}, {Muzahid}, \& {Schaye}}]{ahoranta2021}
{Ahoranta} J., {Finoguenov} A., {Bonamente} M., {Tilton} E., {Wijers} N.,
  {Muzahid} S., {Schaye} J., 2021, \aap, 656, A107

\bibitem[{{Ahoranta} {et~al}\mbox{.}(2020){Ahoranta}, {Nevalainen}, {Wijers},
  {Finoguenov}, {Bonamente}, {Tempel}, {Tilton}, {Schaye}, {Kaastra}, \&
  {Gozaliasl}}]{ahoranta2020}
{Ahoranta} J. {et~al.}, 2020, \aap, 634, A106

\bibitem[{Anderson \& Darling(1952)}]{anderson1952}
Anderson T.~W., Darling D.~A., 1952, The Annals of Mathematical Statistics, 23,
  193

\bibitem[{{Arnaud}(1996)}]{arnaud1996}
{Arnaud} K.~A., 1996, in Astr. Data Analysis Software and Systems V, {Jacoby}
  G.~H., {Barnes} J., eds., Vol. 101, p.~17

\bibitem[{{Baker} \& {Cousins}(1984)}]{cousins1984}
{Baker} S., {Cousins} R.~D., 1984, Nuclear Instruments and Methods in Physics
  Research, 221, 437

\bibitem[{{Bevington} \& {Robinson}(2003)}]{bevington2003}
{Bevington} P.~R., {Robinson} D.~K., 2003, Data reduction and error analysis
  for the physical sciences. McGraw Hill, Third Edition

\bibitem[{Bishop {et~al}\mbox{.}(1975)Bishop, Fienberg, \&
  Holland}]{bishop1975}
Bishop Y., Fienberg S., Holland P., 1975, Discrete Multivariate Analysis:
  Theory and Practice : Yvonne M.M. Bishop, Stephen E. Fienberg and Paul W.
  Holland. Massachusetts Institute of Technology Press

\bibitem[{Bonamente(2019)}]{bonamente2019}
Bonamente M., 2019, Journal of Applied Statistics, 46, 1129

\bibitem[{Bonamente(2020)}]{bonamente2020}
Bonamente M., 2020, Journal of Applied Statistics, 47, 2044

\bibitem[{{Bonamente}(2022)}]{bonamente2022book}
{Bonamente} M., 2022, Statistics and Analysis of Scientific Data. Springer,
  Graduate Texts in Physics, Third Edition

\bibitem[{{Bonamente} {et~al}\mbox{.}(2003){Bonamente}, {Joy}, \&
  {Lieu}}]{bonamente2003}
{Bonamente} M., {Joy} M.~K., {Lieu} R., 2003, \apj, 585, 722

\bibitem[{{Bonamente} {et~al}\mbox{.}(2016){Bonamente}, {Nevalainen}, {Tilton},
  {Liivam{\"a}gi}, {Tempel}, {Hein{\"a}m{\"a}ki}, \& {Fang}}]{bonamente2016}
{Bonamente} M., {Nevalainen} J., {Tilton} E., {Liivam{\"a}gi} J., {Tempel} E.,
  {Hein{\"a}m{\"a}ki} P., {Fang} T., 2016, \mnras, 457, 4236

\bibitem[{Bonamente \& Spence(2022)}]{bonamente2022}
Bonamente M., Spence D., 2022, Journal of Applied Statistics, 49, 522

\bibitem[{Cameron \& Trivedi(1986)}]{cameron1986}
Cameron A.~C., Trivedi P.~K., 1986, Journal of Applied Econometrics, 1, 29

\bibitem[{{Cameron} \& {Trivedi}(2013)}]{cameron2013}
{Cameron} C., {Trivedi} P., 2013, Regression Analysis of Count Data (Second
  Ed.). Cambridge University Press

\bibitem[{{Cash}(1976)}]{cash1976}
{Cash} W., 1976, \aap, 52, 307

\bibitem[{{Cash}(1979)}]{cash1979}
{Cash} W., 1979, \apj, 228, 939

\bibitem[{Consul \& Jain(1973)}]{consul1973}
Consul P.~C., Jain G.~C., 1973, Technometrics, 15, 791

\bibitem[{Conway \& Maxwell(1962)}]{conway1962}
Conway T.~R., Maxwell W.~L., 1962, Journal of Industrial Engineering, 12, 132

\bibitem[{Cramer(1946)}]{cramer1946}
Cramer H., 1946, Mathematical Methods of Statistics. Princeton University
  Press, Princeton

\bibitem[{Dean \& Lawless(1989)}]{dean1989}
Dean C., Lawless J.~F., 1989, Journal of the American Statistical Association,
  84, 467

\bibitem[{Dean(1992)}]{dean1992}
Dean C.~B., 1992, Journal of the American Statistical Association, 87, 451

\bibitem[{Famoye(1993)}]{famoye1993}
Famoye F., 1993, Communications in Statistics - Theory and Methods, 22, 1335

\bibitem[{{Fang} {et~al}\mbox{.}(2010){Fang}, {Buote}, {Humphrey}, {Canizares},
  {Zappacosta}, {Maiolino}, {Tagliaferri}, \& {Gastaldello}}]{fang2010}
{Fang} T., {Buote} D.~A., {Humphrey} P.~J., {Canizares} C.~R., {Zappacosta} L.,
  {Maiolino} R., {Tagliaferri} G., {Gastaldello} F., 2010, \apj, 714, 1715

\bibitem[{{Fang} {et~al}\mbox{.}(2007){Fang}, {Canizares}, \& {Yao}}]{fang2007}
{Fang} T., {Canizares} C.~R., {Yao} Y., 2007, \apj, 670, 992

\bibitem[{{Fang} {et~al}\mbox{.}(2002){Fang}, {Marshall}, {Lee}, {Davis}, \&
  {Canizares}}]{fang2002}
{Fang} T., {Marshall} H.~L., {Lee} J.~C., {Davis} D.~S., {Canizares} C.~R.,
  2002, \apjl, 572, L127

\bibitem[{Fisher(1922)}]{fisher1922}
Fisher R., 1922, Journal of the Royal Statistical Society, 85, 87

\bibitem[{Fisher(1934)}]{fisher1934}
Fisher R., 1934, Statistical Methods for Research Workers, Fifth Ed. Oliver and
  Boyd, Edinburgh

\bibitem[{{Giacconi} {et~al}\mbox{.}(1971){Giacconi}, {Kellogg}, {Gorenstein},
  {Gursky}, \& {Tananbaum}}]{giacconi1971}
{Giacconi} R., {Kellogg} E., {Gorenstein} P., {Gursky} H., {Tananbaum} H.,
  1971, \apjl, 165, L27

\bibitem[{Gourieroux {et~al}\mbox{.}(1984{\natexlab{a}})Gourieroux, Monfort, \&
  Trognon}]{gourieroux1984b}
Gourieroux C., Monfort A., Trognon A., 1984{\natexlab{a}}, Econometrica, 52,
  701

\bibitem[{Gourieroux {et~al}\mbox{.}(1984{\natexlab{b}})Gourieroux, Monfort, \&
  Trognon}]{gourieroux1984}
Gourieroux C., Monfort A., Trognon A., 1984{\natexlab{b}}, Econometrica, 52,
  681

\bibitem[{{Greenwood} \& {Nikulin}(1996)}]{greenwood1996}
{Greenwood} P., {Nikulin} M., 1996, A Guide to Chi-Squared Testing. Wiley

\bibitem[{Hilbe(2011)}]{hilbe2011}
Hilbe J.~M., 2011, Negative Binomial Regression, 2nd edn. Cambridge University
  Press

\bibitem[{Hilbe(2014)}]{hilbe2014}
Hilbe J.~M., 2014, Modeling Count Data. Cambridge University Press

\bibitem[{{Humphrey} {et~al}\mbox{.}(2009){Humphrey}, {Liu}, \&
  {Buote}}]{humphrey2009}
{Humphrey} P.~J., {Liu} W., {Buote} D.~A., 2009, \apj, 693, 822

\bibitem[{{Kaastra}(2017)}]{kaastra2017}
{Kaastra} J.~S., 2017, Astronomy and Astrophysics, 605, A51

\bibitem[{{Kaastra} {et~al}\mbox{.}(1996){Kaastra}, {Mewe}, \&
  {Nieuwenhuijzen}}]{kaastra1996}
{Kaastra} J.~S., {Mewe} R., {Nieuwenhuijzen} H., 1996, in UV and X-ray
  Spectroscopy of Astrophysical and Laboratory Plasmas, {Yamashita} K.,
  {Watanabe} T., eds., pp. 411--414

\bibitem[{{Kaastra} {et~al}\mbox{.}(2006){Kaastra}, {Werner}, {Herder},
  {Paerels}, {de Plaa}, {Rasmussen}, \& {de Vries}}]{kaastra2006}
{Kaastra} J.~S., {Werner} N., {Herder} J.~W.~A.~d., {Paerels} F.~B.~S., {de
  Plaa} J., {Rasmussen} A.~P., {de Vries} C.~P., 2006, \apj, 652, 189

\bibitem[{Kolmogorov(1933)}]{kolmogorov1933}
Kolmogorov A., 1933, Giornale dell' Istituto Italiano degli Attuari, 4, 1

\bibitem[{{Kov{\'a}cs} {et~al}\mbox{.}(2019){Kov{\'a}cs}, {Bogd{\'a}n},
  {Smith}, {Kraft}, \& {Forman}}]{kovacs2019}
{Kov{\'a}cs} O.~E., {Bogd{\'a}n} {\'A}., {Smith} R.~K., {Kraft} R.~P., {Forman}
  W.~R., 2019, \apj, 872, 83

\bibitem[{{Lampton} {et~al}\mbox{.}(1976){Lampton}, {Margon}, \&
  {Bowyer}}]{lampton1976}
{Lampton} M., {Margon} B., {Bowyer} S., 1976, \apj, 208, 177

\bibitem[{{Lee} {et~al}\mbox{.}(2011){Lee}, {Kashyap}, {van Dyk}, {Connors},
  {Drake}, {Izem}, {Meng}, {Min}, {Park}, {Ratzlaff}, {Siemiginowska}, \&
  {Zezas}}]{lee2011}
{Lee} H. {et~al.}, 2011, \apj, 731, 126

\bibitem[{{Marshall} {et~al}\mbox{.}(2021){Marshall}, {Chen}, {Drake},
  {Guainazzi}, {Kashyap}, {Meng}, {Plucinsky}, {Ratzlaff}, {van Dyk}, \&
  {Wang}}]{marshall2021}
{Marshall} H.~L. {et~al.}, 2021, \aj, 162, 254

\bibitem[{McCullagh(1984)}]{mccullagh1984}
McCullagh P., 1984, Biometrika, 71, 461

\bibitem[{McCullagh(1986)}]{mccullagh1986}
McCullagh P., 1986, Journal of the American Statistical Association, 81, 104

\bibitem[{{McCullagh} \& {Nelder}(1989)}]{mccullagh1989}
{McCullagh} P., {Nelder} J., 1989, Generalized Linear Models. Chapman \&
  Hall/CRC, Second Edition

\bibitem[{{Nevalainen} {et~al}\mbox{.}(2019){Nevalainen}, {Tempel}, {Ahoranta},
  {Liivam{\"a}gi}, {Bonamente}, {Tilton}, {Kaastra}, {Fang},
  {Hein{\"a}m{\"a}ki}, {Saar}, \& {Finoguenov}}]{nevalainen2019}
{Nevalainen} J. {et~al.}, 2019, \aap, 621, A88

\bibitem[{{Nicastro} {et~al}\mbox{.}(2018){Nicastro}, {Kaastra}, {Krongold},
  {Borgani}, {Branchini}, {Cen}, {Dadina}, {Danforth}, {Elvis}, {Fiore},
  {Gupta}, {Mathur}, {Mayya}, {Paerels}, {Piro}, {Rosa-Gonzalez}, {Schaye},
  {Shull}, {Torres-Zafra}, {Wijers}, \& {Zappacosta}}]{nicastro2018}
{Nicastro} F. {et~al.}, 2018, Nature, 558, 406

\bibitem[{{Nicastro} {et~al}\mbox{.}(2005){Nicastro}, {Mathur}, {Elvis},
  {Drake}, {Fiore}, {Fang}, {Fruscione}, {Krongold}, {Marshall}, \&
  {Williams}}]{nicastro2005}
{Nicastro} F. {et~al.}, 2005, \apj, 629, 700

\bibitem[{{Protassov} {et~al}\mbox{.}(2002){Protassov}, {van Dyk}, {Connors},
  {Kashyap}, \& {Siemiginowska}}]{protassov2002}
{Protassov} R., {van Dyk} D.~A., {Connors} A., {Kashyap} V.~L., {Siemiginowska}
  A., 2002, \apj, 571, 545

\bibitem[{{Ren} {et~al}\mbox{.}(2014){Ren}, {Fang}, \& {Buote}}]{ren2014}
{Ren} B., {Fang} T., {Buote} D.~A., 2014, \apjl, 782, L6

\bibitem[{{Rothschild} {et~al}\mbox{.}(1979){Rothschild}, {Boldt}, {Holt},
  {Serlemitsos}, {Garmire}, {Agrawal}, {Riegler}, {Bowyer}, \&
  {Lampton}}]{rothschild1979}
{Rothschild} R. {et~al.}, 1979, Space Science Instrumentation, 4, 269

\bibitem[{Sellers \& Shmueli(2010)}]{sellers2010}
Sellers K.~F., Shmueli G., 2010, The Annals of Applied Statistics, 4, 943

\bibitem[{Shmueli {et~al}\mbox{.}(2005)Shmueli, Minka, Kadane, Borle, \&
  Boatwright}]{shmueli2005}
Shmueli G., Minka T.~P., Kadane J.~B., Borle S., Boatwright P., 2005, Journal
  of the Royal Statistical Society. Series C (Applied Statistics), 54, 127

\bibitem[{{Spence} {et~al}\mbox{.}(2023){Spence}, {Bonamente}, {Nevalainen},
  {Tuominen}, {Ahoranta}, {Liu}, \& {Wijers}}]{spence2023}
{Spence} D., {Bonamente} M., {Nevalainen} J., {Tuominen} T., {Ahoranta} J.,
  {Liu} W., {Wijers} N., 2023, MNRAS submitted, 0

\bibitem[{Stephens(1974)}]{stephens1974}
Stephens M.~A., 1974, Journal of the American Statistical Association, 69, 730

\bibitem[{Wasserstein \& Lazar(2016)}]{asa2016}
Wasserstein R.~L., Lazar N.~A., 2016, The American Statistician, 70, 129

\bibitem[{Wilks(1938)}]{wilks1938}
Wilks S.~S., 1938, Ann. Math. Statist., 9, 60

\bibitem[{{Xu} {et~al}\mbox{.}(2014){Xu}, {van Dyk}, {Kashyap},
  {Siemiginowska}, {Connors}, {Drake}, {Meng}, {Ratzlaff}, \& {Yu}}]{xu2014}
{Xu} J. {et~al.}, 2014, \apj, 794, 97

\end{thebibliography}


\bsp	
\label{lastpage}
\end{document}